\documentclass[%
reprint,
superscriptaddress,
amsmath,amssymb,
aps,
prl,
floatfix
]{revtex4-2}

\usepackage{booktabs}
\usepackage[T1]{fontenc}
\usepackage[utf8]{inputenc}
\usepackage{bbold}
\usepackage{amsmath}
\usepackage{amssymb}
\usepackage{graphicx}
\usepackage{subscript}
\usepackage[unicode=true,pdfusetitle,
bookmarks=true,bookmarksnumbered=false,bookmarksopen=false,
breaklinks=false,pdfborder={0 0 1},backref=false,colorlinks=false]
{hyperref}
\usepackage{xcolor}
\usepackage[title]{appendix}
\usepackage{bm}
\usepackage{physics}
\usepackage[version=4]{mhchem}
\usepackage{textcomp}
\usepackage{indentfirst}
\usepackage{listings}
\usepackage{svg}
\usepackage[export]{adjustbox}
\usepackage{lipsum} 
\usepackage{multirow}
\usepackage{comment}
\usepackage{amsthm}
\usepackage{here}
\usepackage[normalem]{ulem}
\usepackage{soul} 
\usepackage{wrapfig}
\usepackage[most]{tcolorbox}
\usepackage{tikz}
\usepackage{makecell}
\usepackage{array}
\usepackage{tabularray}

\newcommand{\lyxmathsym}[1]{\ifmmode\begingroup\def\b@ld{bold}
	\text{\ifx\math@version\b@ld\bfseries\fi#1}\endgroup\else#1\fi}

\begin{document}
\title{Magnons and excitons from Bethe-Salpeter equation  and its application to graphene-based magnets}

\author{Simran Chourasia}
\email{simran.chourasia@uam.es, corresponding author}
\affiliation{Departamento de F\'isica de la Materia Condensada, Universidad Aut\'{o}noma de Madrid, E-28049 Madrid, Spain}

\author{Maur\'icio F. C. Martins Quintela}
\email{mfcmquintela@gmail.com, corresponding author}
\affiliation{Departamento de F\'isica de la Materia Condensada, Universidad Aut\'{o}noma de Madrid, E-28049 Madrid, Spain}
\affiliation{Condensed Matter Physics Center (IFIMAC), Universidad Aut\'{o}noma de Madrid, E-28049 Madrid, Spain}

\author{J. J. Palacios}
\affiliation{Departamento de F\'isica de la Materia Condensada, Universidad Aut\'{o}noma de Madrid, E-28049 Madrid, Spain}
\affiliation{Condensed Matter Physics Center (IFIMAC), Universidad Aut\'{o}noma de Madrid, E-28049 Madrid, Spain}
\affiliation{Instituto Nicol\'as Cabrera (INC), Universidad Aut\'onoma de Madrid, E-28049 Madrid, Spain}

\begin{abstract}
	With the recent isolation of two-dimensional semiconducting magnetic materials, the theoretical study of spin-flip collective excitations (magnons) along with spin-conserving electron-hole bound pairs (excitons) has become particularly relevant. Describing both types of quasiparticles within the same theoretical framework is possible, in principle, but the well-established many-body formalism for the calculation of excitons, based on the intertwined Bethe-Salpeter and Hedin's equations, is not satisfactory for magnons, since the usual approximations produce a spurious gapped Goldstone mode. Here, we  show that  excitons and magnons can be computed  on the same footing under the same Bethe-Salpeter formalism, while respecting the gapless Goldstone mode for magnons.  We demonstrate how this can be achieved through the evaluation of the full Bethe-Salpeter equation, beyond the Tamm-Dancoff approximation in model Hamiltonians. As an example, we apply it to ferromagnetic and antiferromagnetic graphene, faithfully represented by the extended Hubbard model. Our results clarify the roles of on-site and non-local interactions in exciton and magnon appearance and identify consistency requirements relevant to extending the approach to first-principles calculations.
\end{abstract}

\maketitle

\section{Introduction\label{sec:intro}}

With the recent isolation of two-dimensional semiconducting magnetic materials \cite{doi:10.1126/sciadv.abq7650,shi_review_2024,phillips2026}, such as $\ce{CrI3}$ \cite{Huang2017}, $\ce{Cr2Ge2Te6}$ \cite{Gong2017}, $\ce{CrSBr}$ \cite{lee2021magnetic}, or $\ce{PtSe2}$ \cite{Avsar2020}, the study of spin-flip excitations (magnons) along with spin-conserving excitations (excitons) has become a focus of increasing interest\cite{Adak2026}.
In these same semiconducting materials, as is also the case in non-magnetic ones, it has been recognized that excitons are key to understanding their sub-gap optical response \cite{PhysRevLett.105.136805,PhysRevB.90.205422}. The coexistence of excitons and magnons, along with their reciprocal couplings, has raised fundamental questions that are being addressed by experimental\cite{sun2025,iakovlev2025,Datta2025} and theoretical\cite{ebrahimian2025,pnl1-7vcl} works. The effects of the presence of excitons and magnons in a system have been discussed with respect to transport phenomena\cite{iakovlev2025,Dirnberger2026}, second-harmonic generation\cite{PhysRevResearch.5.L042032}, negative refraction \cite{Ma2026}, as well as the magnetic control of exciton–polariton  condensation\cite{zhang2026magneticcontrolexcitonpolaritoncondensate} and the magnetic binding of excitons in Mott insulators\cite{Mehio2023}. 

Theoretical and computational studies of excitons in two-dimensional semiconductors are gaining momentum with new implementations and the development of codes such as BerkeleyGW\cite{DESLIPPE20121269}, Yambo\cite{Sangalli_2019}, EDUS\cite{doi:10.1021/acs.jctc.2c00674}, XATU\cite{uria-alvarez_efficient_2024,Garcia.PhysRevResearch.7.013156}, Exciting\cite{doi.org/10.1002/advs.76167}, among others. The methodologies considered can be essentially divided into those using real-time dynamical approaches \cite{doi:10.1021/acs.jctc.2c00674,Jensen.PhysRevA.109.063104,Tumakov.ppgw-myxb} and those based on solving the Bethe-Salpeter equation (BSE) in the frequency domain\cite{DESLIPPE20121269,uria-alvarez_efficient_2024,Garcia.PhysRevResearch.7.013156,Sangalli_2019,doi.org/10.1002/advs.76167,Blase.10.1021/acs.jpclett.0c01875}. Regarding the computation of magnons, theoretical works in the current literature are mainly based on Heisenberg models\cite{10.1021/acs.nanolett.2c02863,ESTERAS2023100072} through linear spin wave theory\cite{Toth_2015} with exchange parameters obtained from density functional theory (DFT) calculations\cite{Durhuus_2023,PhysRevB.110.045423,t7tm-nxyl,amini2025}. The computation of the parameters has been standardized by codes such as TB2J\cite{HE2021107938}. On the other hand, many efforts have been reported to calculate magnons from time-dependent density functional theory (TDDFT)\cite{Buczek2009,Savrasov1998,BuczekErnstSandratskii2011b,BuczekErnstBrunoSandratskii2009,LounisCostaMunizMills2010,LounisCostaMunizMills2011,LounisDosSantosDiasSchweflinghaus2015,DosSantosDiasSchweflinghausBluegelLounis2015,RousseauEigurenBergara2012,Skovhus.PhysRevB.106.085131,Skovhus.PhysRevB.103.245110,gorni2018spin,tancogne2020time_arxiv} or many-body perturbation theory\cite{Muller2016AcousticMagnons, KarlssonAryasetiawan2000,KotanivanSchilfgaarde2008, Sasioglu2010,489b-n5pd,Olsen.PhysRevLett.127.166402,Ke2021}.
These methods have been mostly confined to the study of standard ferromagnets, but their use has recently been extended to 2D magnets \cite{gbpw-zh1v,Olsen.PhysRevLett.127.166402,Ke2021,Beida2025}. 



One of the simplest and yet still experimentally relevant systems is that of hydrogenated graphene \cite{Yazyev_2010,Han2014,PhysRevB.81.165409}, where the controlled adsorption of hydrogen atoms creates unpaired electrons which can couple ferro or antiferromagnetically  \cite{doi:10.1126/science.aaf4521,TJUNG201797,doi:10.1126/science.aad8038}. Recently, it has been shown that graphene can host all nonrelativistic magnetic phases (\emph{i.e.}, diamagnetism, ferromagnetism, antiferromagnetism, paramagnetism, altermagnetism, ferrimagnetism, and fully compensated ferrimagnetism)\cite{vina-bausa_building_2025}. 
In this work, we are interested in the evaluation of excitons and magnons in this system using the BSE as a common methodology. By focusing on model Hamiltonians, we single out one of the reasons for a known failure of the  application of the BSE formalism to magnons, the appearance of a gap in the Goldstone mode, and propose how to successfully address this problem. Our conclusions logically extend to the first-principles evaluation of magnons under the BSE, although there are other reasons for this more challenging case that are still being discussed in the literature\cite{Muller2016AcousticMagnons}.

We begin in Sec. \ref{overviewBSE} by providing an overview of the BSE in its various forms and approximations.  
In Sec. \ref{sec:sec3}, we narrow the discussion down to model Hamiltonians, in particular focusing on the extended Hubbard model. We  provide explicit expressions for the BSE, dissecting the kernel according to the different interaction terms in the Hamiltonian. We also clarify why the (gapless) Goldstone can be obtained within the BSE formalism.   
Finally, in Sec. \ref{sec:sec4}, we apply the previously discussed methodology to the computation of excitons and magnons in magnetic graphene (ferromagnetic and antiferromagnetic), for which the  extended Hubbard Hamiltonian provides a faithful description.

\section{different forms of the Bethe-Salpeter equation \label{overviewBSE}}

Consider a generic Hamiltonian written as
\begin{equation}
    H=H^{\rm MF}+(V_{\rm e-e}-V^{\rm  MF}_{\rm e-e}),
\label{Hamiltonian}
\end{equation}
where the first term
$$
H^{\rm MF}=H_0+V_{e-e}^{\rm MF}
$$
is a self-consistent mean-field (MF) Hamiltonian with ${H}_0$ the non-interacting term (kinetic plus electron-ion interaction energy) and $V_{e-e}^{\rm MF}$ the MF potential. The second term in Eq.~\ref{Hamiltonian} thus represents the ``residual'' interaction. Our main focus here is on non-relativistic systems, so spin-orbit coupling will be disregarded throughout this work. We denote the creation and annihilation operators of a Bloch electron in a state $\psi_{n \boldsymbol{k} \sigma}(\boldsymbol{r})$ with band index $n$, crystal momentum $\bm k$, spin $\sigma$, and energy $\xi_{n \boldsymbol{k} \sigma}$  by $c_{n \boldsymbol{k} \sigma}^{\dagger}$ and $c_{n \boldsymbol{k} \sigma}$, respectively, where these operators diagonalize the MF Hamiltonian: 
$$
{H}^{\rm MF}=\sum_{n \boldsymbol{k} \sigma} \xi_{n \boldsymbol{k} \sigma} c_{n \boldsymbol{k} \sigma}^{\dagger} c_{n \boldsymbol{k} \sigma}.
$$
In this basis, the term  $V_{\rm e-e}$ is given by 
$$
\begin{aligned}
&{V}_{\mathrm{e}-\mathrm{e}}= \\
&\frac{1}{2} \sum_{\substack{\left\{n_i,\boldsymbol{k}_i\right\} \\
\sigma, \sigma^{\prime}}} V_{n \bm{k_1}\sigma, m \bm{k_2}\sigma'; m' \bm{k_4}\sigma, n' \bm{k_3}\sigma'} c_{n \boldsymbol{k}_1 \sigma}^{\dagger} c_{m \boldsymbol{k}_2 \sigma^{\prime}}^{\dagger} c_{n^\prime \boldsymbol{k}_3 \sigma^{\prime}} c_{m^\prime \boldsymbol{k}_4 \sigma}
\end{aligned}
$$
with 
\begin{align}
	& V_{n \bm{k_1}\sigma, m \bm{k_2}\sigma'; m' \bm{k_3}\sigma, n' \bm{k_4}\sigma'} = \nonumber \\
	&\int d\bm{r} d\bm{r'} \psi_{n \bm{k_1}\sigma}^*(\bm{r})\psi_{m \bm{k_2}\sigma'}^*(\bm{r'})V(\bm{r},\bm{r'})\psi_{m' \bm{k_3}\sigma}(\bm{r})\psi_{n' \bm{k_4}\sigma'} (\bm{r'}).\label{eq:integral_interaction}
\end{align}
In principle, the bare interaction should be considered here. In model Hamiltonians, as shown below, simplifications are made for both the non-interacting and interaction terms.

Both excitons, considered here as spin-conserving electron-hole (e-h) bound pairs, and magnons, spin-flip collective excitations, can, in principle, be evaluated  solving the BSE in its various forms. The bound nature of the electron and the hole and an associated energy below the band gap is characteristic of the exciton, although the BSE can also describe unbound e-h  pairs in the continuum.  On the other hand, we will generally refer to a magnon when the excitation flips a single spin from  majority bands to  minority bands and lowers the total spin (total $S_z$ component) by one unit. Furthermore, we will refer to magnon only when the character of this excitation is collective rather than particle-hole type. The latter are known as Stoner-like excitations and can also be obtained from the BSE. Spin-flip excitations from minority bands to majority bands, that is, those raising the total spin by one unit {in the ferromagnetic system} will be simply called dark excitons, regardless of its binding nature.

Here,  we briefly summarize the different approaches and fix our terminology. 
{Notice that it feeds on nuclear physics, quantum chemistry, and condensed matter terminologies:
\begin{itemize}
\item Independent particle approximation (IPA). The ground state of the system is considered to be a single Slater determinant, $\ket{\Phi_0}$, obtained from a MF procedure, either Hartree-Fock (HF) or DFT. The excitations are approximated by single particle-hole pairs:
$$\ket{X}^+=c_{c\bm k\sigma^\prime}^\dagger c_{v\bm k\sigma}\ket{\Phi_0},$$
where $c^\dagger/c$ create/destroy MF quasiparticles with $c,v$ being conduction and valence bands. Clearly, $\ket{X}^-=c_{v\bm k\sigma^\prime}^\dagger c_{c\bm k\sigma}\ket{\Phi_0}=0$. This approximation is only valid if the two-body residual interactions are neglected in the Hamiltonian and may be unnecessarily labelled as BSE/IPA.
\item Tamm-Dancoff approximation (TDA). The ground state is also approximated by a MF single Slater determinant $\ket{\Phi_0}$, but the excitations are now collective e-h pairs:
\begin{equation}
\ket{X}^+ = \sum_{c,v,\bm k} A^{\sigma'\sigma}_{c,v}(\bm k) c_{c\bm k\sigma^\prime}^\dagger c_{v\bm k\sigma}\ket{\Phi_0}.
\label{eq:exciton}
\end{equation}
Again, $\ket{X}^-=\sum_{c,v,\bm k} B_{c,v}^{\sigma'\sigma}(\bm k)c_{v\bm k\sigma^\prime}^\dagger c_{c\bm k\sigma}\ket{\Phi_0}=0$. Choosing HF as the MF approximation, one obtains the BSE/TDA equation through equation of motion (EOM), although a similar result can be obtained by applying the variational principle to the excitation or through a simplification of the linearized time-dependent HF (TDHF) equations. 
In practice, however, the MF ground state is preferably obtained within a DFT approximation. The  quasi-particle energies resulting in the BSE/TDA equations are then substituted by the DFT ones  and the HF kernel by the Hartree plus exchange correlation kernel (this is actually justified as an approximation to time-dependent DFT (TDDFT)). Alternatively, an effective screened interaction can be considered in the HF kernel, while still using the DFT quasiparticle energies\cite{uria-alvarez_efficient_2024}.   If the starting point is a model Hamiltonian, these considerations do not apply and the MF plus the BSE/TDA calculations are carried out  simply on the basis of the model Hamiltonian at the HF level. 
\item Random-phase approximation (RPA).  The ground state $\ket{\tilde\Phi_0}$ is no longer a single Slater determinant, being a correlated state composed of an infinite number of configurations with even-numbered e-h pair excitations\cite{THOULESS1960225}.  The collective excitations 
are now considered to be
\begin{equation}
    \quad\ket{X} = \sum_{c,v,\bm k} \left( A_{c,v}^{\sigma'\sigma}(\bm k) c_{c\bm k\sigma^\prime}^\dagger c_{v\bm k\sigma} - B^{\sigma'\sigma}_{c,v}(\bm k) c_{v\bm k\sigma^\prime}^\dagger c_{c\bm k\sigma}\right)\ket{\tilde\Phi_0}.\label{eq:rpa_excitation}
\end{equation}
Again, the EOM leads to the BSE/RPA equation, also known as the Casida equation\cite{Casida.doi:10.1142/9789812830586_0005} or full BSE. The actual ground state $\ket{\tilde\Phi_0}$ is, in practice, not needed, with the derivation still based on $\ket{\Phi_0}$ {using the quasi-boson approximation (QBA) of} EOM~\cite{Co_2023}. {A common approximation in solids involves discarding the Fock term to get a simplified version of the RPA which is still useful to describe plasmons.} A linearized time-dependent HF or TDDFT leads to the same type of equation and similar alternatives as those in the BSE/TDA apply here as well. HF is deprecated in favour of DFT and TDDFT actually justifies the use of DFT quasiparticle energies as in the TDA case. The actual exchange-correlation kernel or, alternatively, a HF screened effective {kernel} can be used. Again, in the case of model Hamiltonians, the bare HF approximation is the natural choice.
 \item BSE/GW approximation.  This fully first-principles approach is based on Green's function many-body perturbation theory where screening is explicitly taken into account through Hedin's equations\cite{Blase.10.1021/acs.jpclett.0c01875}. In its most popular form, the quasiparticle energies are replaced by GW quasiparticle energies, hence the name. In other words, the local and static DFT  self-energy operator is replaced by the non-local and dynamical GW self-energy operator in the calculation of the quasiparticle energies.  The direct term in the kernel is usually screened by a static dielectric matrix.  Both TDA and RPA flavours can be used, {with the TDA the more frequently used due to its relative computational simplicity}.
\end{itemize}}

{ The different flavours of the BSE discussed above apply equally to the discussion of e-h pairs with finite centre-of-mass (COM) momentum  and to the calculation of spin-conserving ($\sigma^\prime=\sigma$) or spin-flip ($\sigma^\prime=\bar{\sigma}$) excitations. However, it is well documented\cite{Muller2016AcousticMagnons,Cooke1973,CookeBlackmanMorgan1985,CookeLynnDavis1980,KarlssonAryasetiawan2000,KotanivanSchilfgaarde2008,Sasioglu2010,Savrasov1998,Buczek2009,BuczekErnstSandratskii2011a,BuczekErnstSandratskii2011b,BuczekErnstBrunoSandratskii2009,LounisCostaMunizMills2010,LounisCostaMunizMills2011,LounisDosSantosDiasSchweflinghaus2015,DosSantosDiasSchweflinghausBluegelLounis2015,RousseauEigurenBergara2012,vanSchilfgaardeKotaniFaleev2006,FaleevvanSchilfgaardeKotani2004} that even the most ambitious calculations to date, namely,  BSE/GW/TDA, cannot reproduce the zero-energy Goldstone mode in magnetic systems, essentially because of the inconsistency between the dynamical screening in the  quasiparticle self-energies and the static screening in the interaction kernel of the BSE. Importantly, we note that fixing this issue does not solve the problem, as we discuss below.}

\section{magnons and excitons in model Hamiltonians\label{sec:sec3}}
{
Without loss of generality, here we are interested in model Hamiltonians such as 
the single-site extended Hubbard model:
\begin{equation}
\begin{aligned}
{H}&={T}+{V}\\
{T}&=\sum_{i,j}t\,c_{i\sigma}^\dagger c_{j\sigma}\\
    {V} &= U \sum_{i} {n}_{i\uparrow} {n}_{i\downarrow} + J \sum_{<i,j>} \sum_{\sigma, \sigma'} {n}_{i\sigma}{n}_{j\sigma'},
 \end{aligned}\label{eq:gen_hubbard}
\end{equation}
where ${c}_{i\sigma}^\dagger ({c}_{i \sigma})$ is the creation (annihilation) operator for an electron with spin $\sigma$ at site $i$, $t$ is the hopping amplitude (only $\neq 0$ for nearest neighbours), ${n}_{i\sigma} = {c}_{i\sigma}^\dagger {c}_{i \sigma}$ is the number density operator, and $<i,j>$ represents that the summation is done over nearest-neighbouring sites with indices $i$ and $j$, each pair counted once.  }

In this case, the natural choice of action to compute excitons and magnons is through BSE/TDA or BSE/RPA, starting from a reference HF ground state. The inconsistency mentioned at the end of the previous section does not appear in this case since the quasi-particle self-energies and the interaction kernel are naturally evaluated using the same interaction model. This singles out a second, frequently overlooked difficulty when it comes to magnon computation, namely, that $S^2$ is typically not a good quantum number in the MF ground state $\ket{\Phi_0}$. This also opens a gap in the Goldstone mode \cite{Cui.10.1063/1.4824905} at the BSE/TDA level. Here, we demonstrate that it is possible to fix this by using BSE/RPA on an unrestricted HF (UHF) reference ground state.

\subsection{BSE/TDA}
The simplest way to arrive at the BSE/TDA is to apply the variational principle to the excitation proposed in Eq. \ref{eq:exciton}. This leads to an eigenvalue equation where
\begin{align} \label{eq:bse-res}
	& H_{vc,v'c'}\left(\bm{k},\bm{k}',\bm{Q}\right) =\bra{v,c, \bm{k}, \bm{Q}} {H} \ket{v',c', \bm{k'}, \bm{Q}}= \nonumber \\     &=\delta_{\bm{k},\bm{k}'}\delta_{vv'}\left[\delta_{cc'} \varepsilon_{c\bm{k} + \bm{Q}} + \Sigma_{cc'}\left(\bm{k}+\bm{Q},\bm{k}+\bm{Q}\right)\right] \nonumber \\
    &\quad -\delta_{\bm{k},\bm{k}'} \delta_{cc'}\left[\delta_{vv'} \varepsilon_{v\bm{k}}
    +\Sigma_{v'v}\left(\bm{k},\bm{k}\right)\right]\nonumber\\
    & \quad -D_{vc,v'c'} \left(\bm{k},\bm{k}',\bm{Q}\right)+X_{vc,v'c'}\left(\bm{k},\bm{k}',\bm{Q}\right),    
\end{align}  
is the Hamiltonian in the single e-h pair excitation basis $\ket{v,c,\bm{k}, \bm{Q}}={c}^\dagger_{c \bm{k+Q}} {c}_{v \bm{k}} \ket{\Phi_0}$ with respect to the MF ground state $\ket{\Phi_0}$. 
$\varepsilon_{c/v\bm{k}}$ stands for non-interacting energies and the quasiparticle self-energy is given by
\begin{align}
	\Sigma_{nm}\left(\bm{k},\bm{k}'\right) =&  \sum_{v'',\bm{k}''}^{\mathrm{occ}}\big(V_{n\bm{k},v''\bm{k}'';m\bm{k}',v''\bm{k}''} \nonumber \\ 
    & -V_{n\bm{k},v''\bm{k}'';v''\bm{k}'',m\bm{k}'}\big),
\end{align}
where the summation runs over all the occupied states. The MF quasi-particle energies $\xi_{c/v\bm k} $ result from diagonalizing
$$
 \varepsilon_{c/v\bm{k}}+ \Sigma_{nm}\left(\bm{k}\right).
$$
The direct term 
\begin{align}\label{eq:direct}	D_{vcv'c'}\left(\bm{k},\bm{k}',\bm{Q}\right)&=V_{c\bm{k}+\bm{Q},v'\bm{k'};c'\bm{k'}+\bm{Q},v\bm{k}},
\end{align}
and the exchange term 
\begin{align}\label{eq:exchange}	X_{vcv'c'}\left(\bm{k},\bm{k}',\bm{Q}\right)&=V_{c\bm{k}+\bm{Q},v'\bm{k'};v\bm{k},c'\bm{k'}+\bm{Q}},
\end{align}
form the BSE kernel
\begin{align}\label{eq:bse_kernel}
	& K_{vcv'c'}\left(\bm{k},\bm{k}',\bm{Q}\right)= \nonumber \\
    & -D_{vcv'c'} \left(\bm{k},\bm{k}',\bm{Q}\right) + X_{vcv'c'}\left(\bm{k},\bm{k}',\bm{Q}\right).
\end{align}
By comparison with the direct and exchange terms in their integral formulations (see Eq.~\ref{eq:integral_interaction}) and considering the $\delta_{\bm{k},\bm{k}'}$ prefactor in the diagonal terms, we can identify each self-energy term as a direct or exchange self-energy and write
\begin{equation}
	\Sigma_{nm}\left(\bm{k},\bm{k}'\right) = \sum_{v'',\bm{k}''}^{\mathrm{occ}}\left(\Sigma_{nm}^{D}\left(\bm{k},\bm{k}'\right) -\Sigma_{nm}^{X}\left(\bm{k},\bm{k}'\right)\right).
\end{equation}
Notice that, for clarity's sake, we have suppressed the spin indices. One can always safely absorb them in the labels $c/v$. 

As an illustrative exercise, we can consider distinguishable or non-identical (NI) particles so that the non-interacting ground state corresponds to a Hartree ground state $\ket{\Phi_0^{\text{NI}}}$ instead of a Slater determinant. In this hypothetical case, the BSE/TDA equation acquires the form (see Appendix~\ref{sec:distinguishable})
\begin{align} \label{eq:TDA_NI}
	& H^{\text{NI}}_{vc,v'c'}\left(\bm{k},\bm{k}',\bm{Q}\right) \nonumber \\     &=\delta_{\bm{k},\bm{k}'}\delta_{vv'}\left[\delta_{cc'} \varepsilon_{c\bm{k}+\bm{Q}}+\Sigma_{cc'}^{\text{NI}}\left(\bm{k}+\bm{Q}, \bm{k}+\bm{Q}\right)\right] \nonumber \\
    & \quad - \delta_{v v'} \delta_{\bm{k},\bm{k}'} \delta_{cc'} \left[\varepsilon_{v\bm{k}}+\Sigma_{vv}^{\text{NI}} \left(\bm{k}, \bm{k}\right) \right]\nonumber\\
    & \quad - \delta_{vv'} \delta_{\bm{k k'}} D_{vcvc'} \left(\bm{k},\bm{k},\bm{Q}\right) \nonumber \\
    & \quad +(1-\delta_{vv'})X_{vcv'c'}\left(\bm{k},\bm{k}',\bm{Q}\right),
\end{align}  
where $\Sigma_{nm}^{\text{NI}}(\textbf{k}, \textbf{k}')$ is the quasiparticle self-energy for non-identical electrons,
\begin{align}
    \Sigma_{nm}^{\text{NI}}(\textbf{k}, \textbf{k}') = & \sum_{v'', \textbf{k}''}^{\text{occ}} \Big[(1-\delta_{nv''} \delta_{\textbf{kk}''})  \nonumber \\
    &  (1-\delta_{mv''} \delta_{\textbf{k}'\textbf{k}''}) V_{n\textbf{k}, v'' \textbf{k}''; m \textbf{k}', v'' \textbf{k}''} \Big],
\end{align}
and it consists of Hartree integrals only. $\Sigma_{vv'}^{\text{NI}}$ is non-zero only when $v=v'$, whereas the Hartree integral for identical electrons can be non-zero even for $v\neq v'$. Notice that the kernel for non-identical electrons has the direct term $D$ only for $v = v'$, and not for any pair of valence bands. $D_{vcvc'}$ cancels the self energy term for the interaction of the electron added in the conduction band $c/c'$ with the electron removed from the valence band $v$ in $\Sigma_{cc'}^{\text{NI}}$. So, the direct term present in $H^{\text{NI}}_{vc,v'c'}$ is not the general direct term like in Eq.~\ref{eq:bse-res}. Additionally, $\Sigma_{v'v}^{\text{NI}} $ is only non-zero for $v=v'$ and the exchange term $X_{vcv'c'}$ for $v=v'$ is not present. 
{This explicitly shows that the exchange interaction, only present for identical particles is behind the exciton formation, since it is the $D$ term in the kernel that binds the electron and the hole.}

\subsection{BSE/RPA\label{sec:btda}}

Including anti-resonant excitations and their coupling with resonant excitations as described in Sec. \ref{overviewBSE} we arrive at the BSE/RPA equation, also known as Casida's equation \cite{10.1021/bk-1996-0628.ch008,PhysRevB.49.4973,Blase.10.1021/acs.jpclett.0c01875}. Similarly to the BSE/TDA case, we have that the general Hamiltonian remains block-diagonal in $\bm{Q}$: 
\begin{equation}
    H^{\mathrm{e-h}}=\left(\begin{array}{cccc}
\text{\fbox{\ensuremath{\bm{Q}_{1}}}} &  &  & 0\\
 & \text{\fbox{\ensuremath{\bm{Q}_{2}}}} &  & \ldots\\
0 &  & \text{\fbox{\ensuremath{\bm{Q}_{3}}}}\\
 & \vdots &  & \ddots
\end{array}\right),
\end{equation}
where each $\text{\fbox{\ensuremath{\bm{Q}_{i}}}}$ block \cite{10.1021/cr0505627,sander_beyond_2015,Blase.10.1021/acs.jpclett.0c01875,Garcia.PhysRevResearch.7.013156} is now given by
\begin{equation}
\boxed{\bm Q_i}=\begin{pmatrix} H^{(r,r)}(\bm Q_i) &  H^{(r,a)}(\bm Q_i)\\[2pt]
- H^{(r,a)}(\bm Q_i)^{\dagger} & - H^{(a,a)}(\bm Q_i)^{*}\end{pmatrix},
\end{equation}
and where ${H}^{(\mathrm{r},\mathrm{r})}$ is the resonant part of the BSE Hamiltonian as considered in the TDA (\emph{i.e.}, creation of e-h pairs). $- H^{(a,a)\,*}$ is the antiresonant part corresponding to the annihilation of e-h pairs and ${H}^{(\mathrm{r},\mathrm{a})}$ involves the coupling between the two distinct sectors \cite{Garcia.PhysRevResearch.7.013156,urialvarez2025}. Having already discussed
\begin{align}
    & H^{(r,r)}(\bm Q_i)\equiv H_{vc,v'c'}(\bm k,\bm k',\bm Q),
\end{align}
in Eq.~\ref{eq:bse-res}, we now move to the anti-resonant and coupling terms. For the anti-resonant block, we have 
\begin{align}
    & H^{(a,a)}_{vc,v'c'}(\bm k,\bm k',\bm Q)=
\delta_{\bm k\bm k'}\delta_{vv'}\big[\delta_{cc'}\varepsilon_{c\bm k}+\Sigma_{cc'}(\bm k,\bm k)\big]\nonumber\\
&\qquad-\delta_{\bm k\bm k'}\delta_{cc'}\big[\delta_{vv'}\varepsilon_{v\bm k+\bm Q}+\Sigma_{v'v}(\bm k+\bm Q,\bm k+\bm Q)\big]\nonumber\\
&\qquad -V_{c\bm k,\,v'\bm k'+\bm Q;\,c'\bm k',\,v\bm k+\bm Q}
+V_{c\bm k,\,v'\bm k'+\bm Q;\,v\bm k+\bm Q,\,c'\bm k'}.
\end{align}
For the coupling terms, following  existing literature \cite{sander_beyond_2015,Garcia.PhysRevResearch.7.013156}, we have
\begin{align}
{H}^{(\mathrm{r},\mathrm{a})}\left(\bm{Q}_{i}\right)=-\tilde{D}^{(\mathrm{r},\mathrm{a})}+\tilde{X}^{(\mathrm{r},\mathrm{a})},
\end{align}
where 
the off-diagonal kernel terms are 
given by
\begin{align}
&\tilde{D}^{(\mathrm{r},\mathrm{a})} \nonumber\\
& \;=\int d\bm r\,d\bm r'\,\psi^*_{c\bm k+\bm Q}(\bm r)\,\psi^*_{c'\bm k'}(\bm r')\,V(\bm r,\bm r')\,\psi_{v'\bm k'+\bm Q}(\bm r)\,\psi_{v\bm k}(\bm r')\nonumber\\
&\; =V_{c\bm k+\bm Q,\,c'\bm k';\,v'\bm k'+\bm Q,\,v\bm k}
\end{align}
and
\begin{align}
&\tilde{X}^{(\mathrm{r},\mathrm{a})} \nonumber\\
& \;=\int d\bm r\,d\bm r'\,\psi^*_{c\bm k+\bm Q}(\bm r)\,\psi^*_{c'\bm k'}(\bm r')\,V(\bm r,\bm r')\,\psi_{v\bm k}(\bm r)\,\psi_{v'\bm k'+\bm Q}(\bm r')\nonumber\\
&\; =V_{c\bm k+\bm Q,\,c'\bm k';\,v\bm k,\,v'\bm k'+\bm Q}.
\end{align}
It can also be seen that, at $\bm{Q}=0$, $ H^{(a,a)}= H^{(r,r)}$ and $ H^{(r,a)}$ is complex symmetric, so that, with $A\equiv H^{(r,r)}$ and $B\equiv H^{(r,a)}$,
\begin{equation}
\boxed{\bm Q_i}=\begin{pmatrix}A & B\\[2pt]
-B^{*} & -A^{*}\end{pmatrix},\label{eq:btda_q}
\end{equation}
the familiar Casida form \cite{10.1021/cr0505627,sander_beyond_2015,Blase.10.1021/acs.jpclett.0c01875,Garcia.PhysRevResearch.7.013156}.

As each $\text{\fbox{\ensuremath{\bm{Q}_{i}}}}$ block is pseudo-Hermitian following the relation \cite{Garcia.PhysRevResearch.7.013156}
\begin{equation}
    \text{\fbox{\ensuremath{\bm{Q}_{i}}}}^\dagger=\eta\,\text{\fbox{\ensuremath{\bm{Q}_{i}}}}\,\eta,
\end{equation}
where $\eta=\mathrm{diag}(I,-I)$, with $I$ the identity matrix with the same dimensions as each ${H}$ block, the eigenvalues of the system are real and come in pairs $\omega(\bm Q)$, $-\omega(-\bm Q)$ (i.e. $\pm\omega$ with time-reversal symmetry) if 
\begin{align}
    \boxed{\bm K}&= \begin{pmatrix} H^{(r,r)}(\bm Q_i) &  H^{(r,a)}(\bm Q_i)\\[2pt]
 H^{(r,a)}(\bm Q_i)^{\dagger} &  H^{(a,a)}(\bm Q_i)^{*}\end{pmatrix}\nonumber\\
&=\eta\,\boxed{\bm Q_i}
\end{align}
is positive definite, which is Thouless’ stability condition for the mean-field state \cite{THOULESS1960225}. If $\boxed{\bm K}$ is positive semidefinite, its null vectors are zero-frequency modes.

\subsection{Goldstone theorem and BSE/RPA\label{sec:goldstone_btda}}

Goldstone's theorem states that, whenever a continuous symmetry of a system is broken, a zero mode appears in the excitation spectrum of said system\cite{goldstone_broken_1962}. The proof follows from \cite{THOULESS1960225,goldstone_broken_1962} by considering a broken symmetry one-body operator 
\begin{equation}
     G = \sum_{pq} g_{pq} c^\dagger _p c_q
\end{equation}
that commutes with the Hamiltonian. Separating the sum into a particle-hole part and particle-particle plus hole-hole parts, we have
\begin{align}
     G &= \underbrace{\sum_{cv} \left(g_{cv} c^\dagger _c c_v + g_{vc} c^\dagger _v c_c\right)}_{ G_1} + \underbrace{\sum_{cc'} g_{cc'} c^\dagger _c c_{c'} + \sum_{vv'} g_{vv'} c^\dagger _v c_{v'}}_{ G_2}.
\end{align}
We have now temporarily suppressed for clarity's sake the momentum indices.
Since $ G $ commutes with the Hamiltonian, one can apply the equation of motion (EOM) formulation \cite{Ring2004-kx} to write 
\begin{align}
    \underbrace{\left\langle \left[ c^\dagger _v c_c,\left[ H,  G_1 \right] \right]\right\rangle_0}_{T_1} + \underbrace{\left\langle \left[ c^\dagger _v c_c,\left[ H,  G_2 \right] \right]\right\rangle_0}_{T_2} = 0,\label{eq:commutator_tot}
\end{align}
where $\left\langle \cdot \right\rangle_0$ is defined as taking the expectation value relative to the HF ground state $\left|\Phi_0\right\rangle$. Focusing on the second term of Eq.~\ref{eq:commutator_tot}, we apply the Jacobi identity to write 
\begin{align}
    T_2&=\left\langle \left[ \left[c^\dagger _v c_c, H\right],  G_2  \right]\right\rangle_0 + \left\langle \left[ H,\left[ c^\dagger _v c_c,  G_2 \right] \right]\right\rangle_0,
\end{align}
which can be further simplified knowing that
\begin{align}
     &c^\dagger _c c_{c'}\left|\Phi_0\right\rangle = 0, & c^\dagger _v c_{v'}\left|\Phi_0\right\rangle = \delta_{vv'}\left|\Phi_0\right\rangle,
\end{align}
meaning 
\begin{align}
    \left\langle \left[ \left[c^\dagger _v c_c, H\right],  G_2  \right]\right\rangle_0 = 0
\end{align}
by applying $ G_2$ to the left and to the right. 
For the commutator between $c^\dagger _v c_c$ and $ G_2$, we can easily see that 
\begin{align}
    \left[ c^\dagger _v c_c,  G_2 \right] &\simeq \left[ c^\dagger _v c_c, c^\dagger_{c'} c_{c''} \right]+\left[ c^\dagger_v c_c, c^\dagger_{v'} c_{v''} \right]\nonumber\\
    &\simeq \delta_{cc'} c^\dagger_v c_{c''} - \delta_{vc''} c^\dagger_{c'} c_c + \delta_{cv'} c^\dagger_v c_{v''} - \delta_{vv''} c^\dagger _{v'} c_c\nonumber\\
    &\simeq\delta_{cc'} c^\dagger_v c_{c''} - \delta_{vv''} c^\dagger _{v'} c_c,
\end{align}
{where we have dropped the coefficients $g$ to focus on the operators themselves,} meaning that the outer commutator then is simply
\begin{align}
    \left\langle \left[ H,\left[ c^\dagger _v c_c,  G_2 \right] \right]\right\rangle_0&\simeq\left\langle \left[ H,c^\dagger _v c_c \right]\right\rangle_0  \nonumber\\
    &\simeq -\bra{\Phi_0}c^\dagger _v c_c H \ket{\Phi_0}\nonumber\\
    &\simeq-\bra{\Phi_{vc}} H \ket{\Phi_0}=0, 
\end{align}
per the {vanishing Hamiltonian matrix elements between the HF ground state and singly-excited determinants} $\ket{\Phi_{vc}}$ (Brillouin's theorem \cite{Szabo1996-qh}). Therefore, we have shown that the $ G_2$ contribution vanishes. 

Regarding $ G_1$, we have that $T_1$ must then also be zero, meaning
\begin{align}
    \left\langle \left[ c^\dagger _v c_c,\left[ H,  G_1 \right] \right]\right\rangle_0 &= \sum_{c'v'}\left(g_{c'v'}\left\langle \left[ c^\dagger _v c_c,\left[ H, c^\dagger_{c'} c_{v'} \right] \right]\right\rangle_0\right.\label{eq:commutator_T1}\\
    &\quad+\left. g_{v'c'}\left\langle \left[ c^\dagger _v c_c,\left[ H, c^\dagger_{v'} c_{c'} \right] \right]\right\rangle_0\right)=0.\nonumber
\end{align}
The two expectation values present are, however, the EOM definitions of the resonant and coupling blocks of Eq.~\ref{eq:btda_q} evaluated in the UHF ground state $\ket{\Phi_0}$ \cite{Ring2004-kx}
\begin{align}
    A_{cv,c'v'} & = \left\langle \left[ c^\dagger _v c_c,\left[ H, c^\dagger_{c'} c_{v'} \right] \right]\right\rangle_0\nonumber\\ 
    B_{cv,c'v'} & = -\left\langle \left[ c^\dagger _v c_c,\left[ H, c^\dagger_{v'} c_{c'} \right] \right]\right\rangle_0 .
\end{align}
This therefore means that Eq.~\ref{eq:commutator_T1} can be written simply as 
\begin{align}
     \sum_{c'v'}\left(g_{c'v'} A_{cv,c'v'} - g_{v'c'} B_{cv,c'v'}\right)=0,\nonumber
\end{align}
which is directly the zero-mode equation for $\boxed{\bm K}$ as 
\begin{align}
    \begin{pmatrix}A & B\\[2pt]
B^{*} & A^{*}\end{pmatrix} \begin{pmatrix}g_{c'v'}\\[2pt]
- g_{v'c'}\end{pmatrix} =\begin{pmatrix}0\\[2pt]
0\end{pmatrix}.\label{eq:nullvector}
\end{align}

Rotations about the magnetization axis leave a collinear state invariant, so the broken symmetry generators are $ S^x$ and $ S^y$\cite{PhysRevLett.108.251602}. Combining them into
\begin{align}
     S^\pm&= S^x\pm i S^y\nonumber\\
	&=\sum_{\substack{nm \\ \bm{k}\bm{k}'\\ \sigma\sigma'}}\bra{n\bm{k}\sigma} S^\pm\ket{m\bm{k}'\sigma'}\, c^\dagger_{n\bm{k}\sigma} c_{m\bm{k}'\sigma'},
\end{align}
where $n,m$ run over the spin-polarized bands of both spins and $\ket{n\bm k \sigma}$ is the single-particle UHF spin-orbital, our null-vectors from Eq.~\ref{eq:nullvector} will then be 
\begin{equation}
    \bm{Z}^\pm_G = \begin{pmatrix}S^{\pm}_{c'v'}\\[2pt]
-S^{\pm}_{v'c'}\end{pmatrix}.
\end{equation}
Writing each UHF spin-orbital as the direct product of a spatial Bloch function and a spin component as
\begin{equation}
    \ket{n\bm{k}\sigma}=\ket{\psi_{n\bm{k}\sigma}}\otimes\ket{\sigma}, 
\end{equation}
we have that 
\begin{align}
     S^\pm&=\sum_{\substack{nm \\ \bm{k}\bm{k}'\\ \sigma\sigma'}}\braket{\psi_{n\bm{k}\sigma}}{\psi_{m\bm{k}'\sigma'}}\bra{\sigma} S^\pm\ket{\sigma'}\, c^\dagger_{n\bm{k}\sigma} c_{m\bm{k}'\sigma'}.
\end{align}
Since $ S^+$ gives the companion zero mode in the $\Delta S_z=+1$ sector, we work with $ S^-$ without loss of generality
\begin{align}
     S^-&=\sum_{\substack{nm \\ \bm{k}\bm{k}'\\ \sigma\sigma'}}\braket{\psi_{n\bm{k}\sigma}}{\psi_{m\bm{k}'\sigma'}}\bra{\sigma} S^-\ket{\sigma'}\, c^\dagger_{n\bm{k}\sigma} c_{m\bm{k}'\sigma'}\nonumber\\
    &=\sum_{\substack{nm \\ \bm{k}\bm{k}'\\ \sigma\sigma'}}\braket{\psi_{n\bm{k}\sigma}}{\psi_{m\bm{k}\sigma'}}\delta_{\bm{k}\bm{k}'}\delta_{\sigma \downarrow}\delta_{\sigma' \uparrow}\, c^\dagger_{n\bm{k}\sigma} c_{m\bm{k}'\sigma'}\nonumber\\
    &=\sum_{\substack{nm \\ \bm{k}}}\braket{\psi_{n\bm{k}\downarrow}}{\psi_{m\bm{k}\uparrow}}\, c^\dagger_{n\bm{k}\downarrow} c_{m\bm{k}\uparrow}\label{eq:matelemS}.
\end{align}
Therefore,
\begin{equation}
    \bm{Z}^-_G = \begin{pmatrix}\braket{\psi_{c\bm{k}\downarrow}}{\psi_{v\bm{k}\uparrow}}\\[2pt]
-\braket{\psi_{v\bm{k}\downarrow}}{\psi_{c\bm{k}\uparrow}}\end{pmatrix}=\begin{pmatrix}\bm{x}_G^-\\[2pt]
\bm{y}_G^-\end{pmatrix}\label{eq:zeromode}
\end{equation}
will be a null-vector of $\boxed{\bm K}$ at $\bm Q=0$ \cite{THOULESS1960225,THOULESS196178}, meaning an exact zero-frequency mode of $\boxed{\bm Q_i}$ (analogously for $S^+$). 
{This requires (i) a self-consistent mean field, (ii) the same interaction in the calculation of the HF ground state and in the BSE kernel (so that the $A$ and $B$ matrices used in the BSE computation match with those stemming from the EOM), and (iii) a complete particle-hole space. }

\begin{figure}
	\centering\includegraphics[scale=0.25]{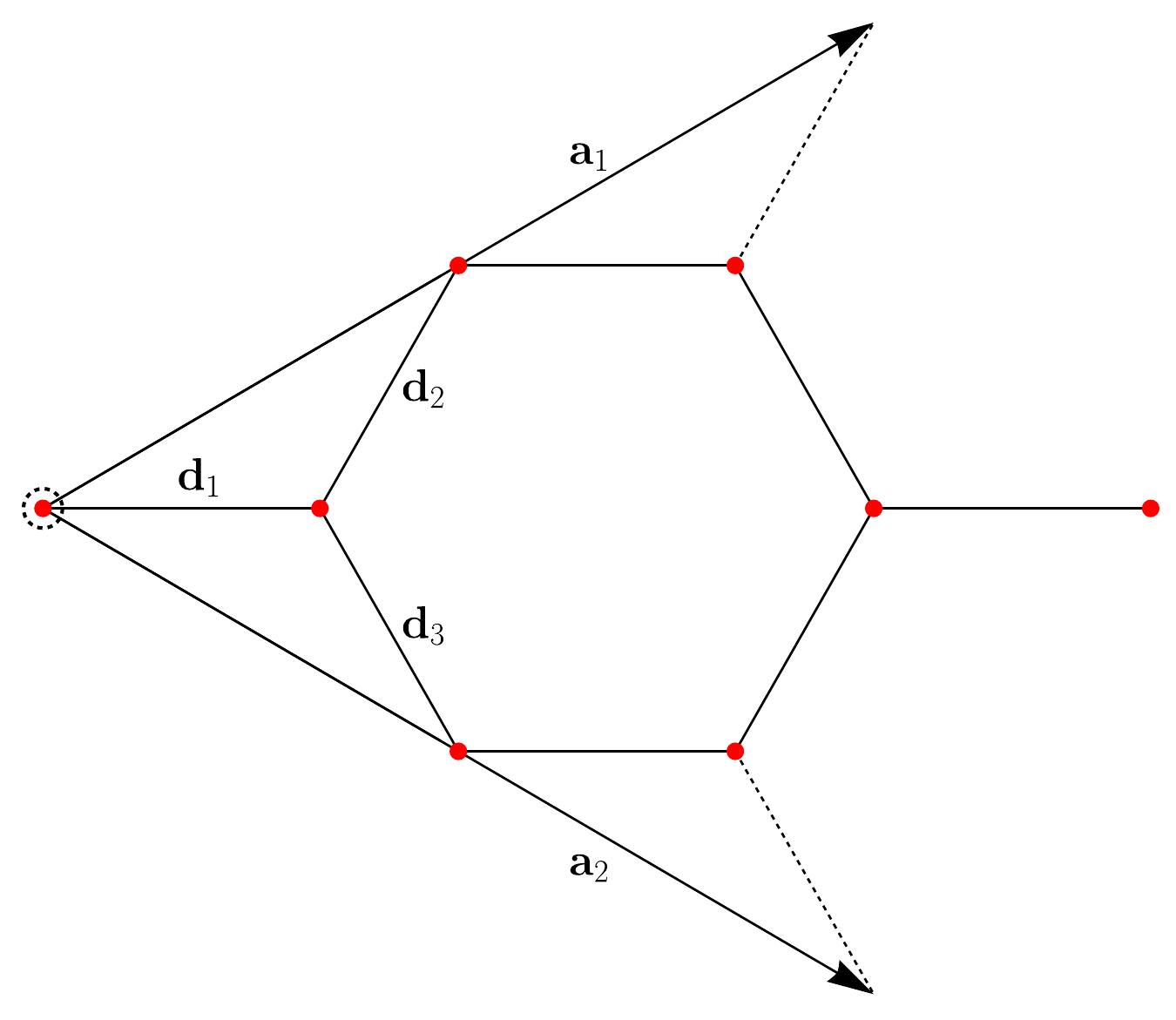}
	\vspace{-0.5cm}\centering\caption{Unit cell considered for hydrogenated graphene system, with $8$ Carbon atoms per unit cell. The dashed circle around the leftmost atomic site indicates the position of the added Hydrogen atom. Lattice vectors $\bm{a}_{1/2}$ correspond to doubling the lattice vectors of the standard graphene lattice.  \label{fig:lattice}}
\end{figure}

Since the BSE/TDA keeps only $ H^{(r,r)}$, it contains the zero mode only if $ H^{(r,r)}\bm{x}_G^-=0$, which requires $\bm{y}_G^-$ to be equivalent to BSE/RPA (or $ H^{(r,a)}$ to be the zero matrix, see Table~\ref{Table:SummaryTable} as to why that is not the case). 
However, for a UHF ground state, $\|\bm{y}_G^-\|^2$ is precisely its spin contamination \cite{Szabo1996-qh,andrews_spin_1991,nandi_effect_1996} as we will now show. Recalling Eq.~\ref{eq:matelemS}, the amplitudes of $ S^+=( S^-)^\dagger$ acting on the ground state are given by 
\begin{equation}
	\begin{aligned}
	 S^+\ket{\Phi_0}=\sum_{\substack{nm \\ \bm{k}}}\braket{\psi_{n\bm{k}\uparrow}}{\psi_{m\bm{k}\downarrow}}\, c^\dagger_{n\bm{k}\uparrow} c_{m\bm{k}\downarrow}\ket{\Phi_0} .
	\end{aligned}
	\label{eq:splus}
\end{equation}
Since $\ket{\Phi_0}$ is an eigenstate of $ S_z$ and $ S^2= S^- S^+ + S_z( S_z+1)$, the norm of Eq.~\ref{eq:splus}, which the orthonormality of the singly excited determinants reduces to a sum of squared overlaps, is the spin contamination as defined in the literature \cite{BAKER1993380,Szabo1996-qh}
\begin{equation}
	\begin{aligned}
	\|\bm{y}_G^-\|^2&=\bra{\Phi_0} S^- S^+ \ket{\Phi_0}\\
	&=\bra{\Phi_0} S^2 -  S_z( S_z+1) \ket{\Phi_0}\\
	&=\langle S^2\rangle_{\rm UHF}-\langle S_z( S_z+1)\rangle_{\rm UHF}\\
    &\equiv\Delta\langle S^2\rangle_{\rm UHF} .
	\end{aligned}
	\label{eq:ynorm}
\end{equation}
Hence, a finite spin contamination means that, in general,  
\begin{equation}
	\Delta\langle S^2\rangle\neq0\Leftrightarrow\bm{y}_G^-\neq0\Leftrightarrow H^{(r,r)}\bm{x}_G^-=- H^{(r,a)}\bm{y}_G^-\neq0,
\end{equation}
so $\bm{x}_G^-$ is not a zero mode of the BSE/TDA and its Goldstone mode is gapped, whereas the BSE/RPA keeps the exact zero. Focusing our attention on spin-flip excitations, it can be seen that exchange type integrals are zero (see Table~\ref{Table:SummaryTable}). In the $\Delta S_z=-1$ sector, $\bm{x}_G^-$ lives on the pairs $(v\!\uparrow,c\!\downarrow)$ and $\bm{y}_G^-$ on $(v\!\downarrow,c\!\uparrow)$, and couple through $\tilde D^{(r,a)}\neq0$  while $\tilde X^{(r,a)}=0$ (see Table~\ref{Table:SummaryTable}). This coupling is what carries $\bm{y}_G^-$ in Eq.~\ref{eq:zeromode} and restores the zero-energy Goldstone mode, as we will see ahead. Remarkably, while a small change is also visible in the higher energy bound magnons, as we will see ahead, working in the BSE/RPA leads to {nearly} identical excitons to {those obtained within the} BSE/TDA {for the systems studied}, as visible in Figs.~\ref{fig:optical-response} and~\ref{fig:magnon_wavefunction}.

The symplectic norm of the zero mode is fixed by the magnetization \cite{10.1063/1.1353585,sander_beyond_2015}
\begin{equation}
	\bm{Z}^{-\dagger}_G\,\eta\,\bm{Z}^-_G=\|\bm{x}_G^-\|^2-\|\bm{y}_G^-\|^2
	=\langle\Phi_0|[ S^+, S^-]|\Phi_0\rangle=2\langle S_z\rangle .
	\label{eq:znorm}
\end{equation}
In a ferromagnet $\langle S_z\rangle\neq0$, giving one Goldstone mode with finite norm and quadratic dispersion (type B) \cite{PhysRev.58.1098}, and the $ S^+$ zero mode has norm $-2\langle S_z\rangle$ and is its $-\omega$ partner.
In an antiferromagnet $\langle S_z\rangle=0$, so the zero mode has zero norm and pairs with its conjugate
coordinate, giving two linear modes (type A) \cite{PhysRev.86.694,PhysRev.87.568}. Both cases satisfy the counting rule $n_A+2n_B=n_{\rm BG}=2$ \cite{PhysRevLett.108.251602,PhysRevLett.110.091601,annurev-conmatphys-031119-050644}, with $n_B=1$ for the ferromagnet and $n_A=2$ for the antiferromagnet.


\section{Ferromagnetic graphene\label{sec:sec4}}

\begin{figure*}
    \centering\includegraphics[scale=0.39]{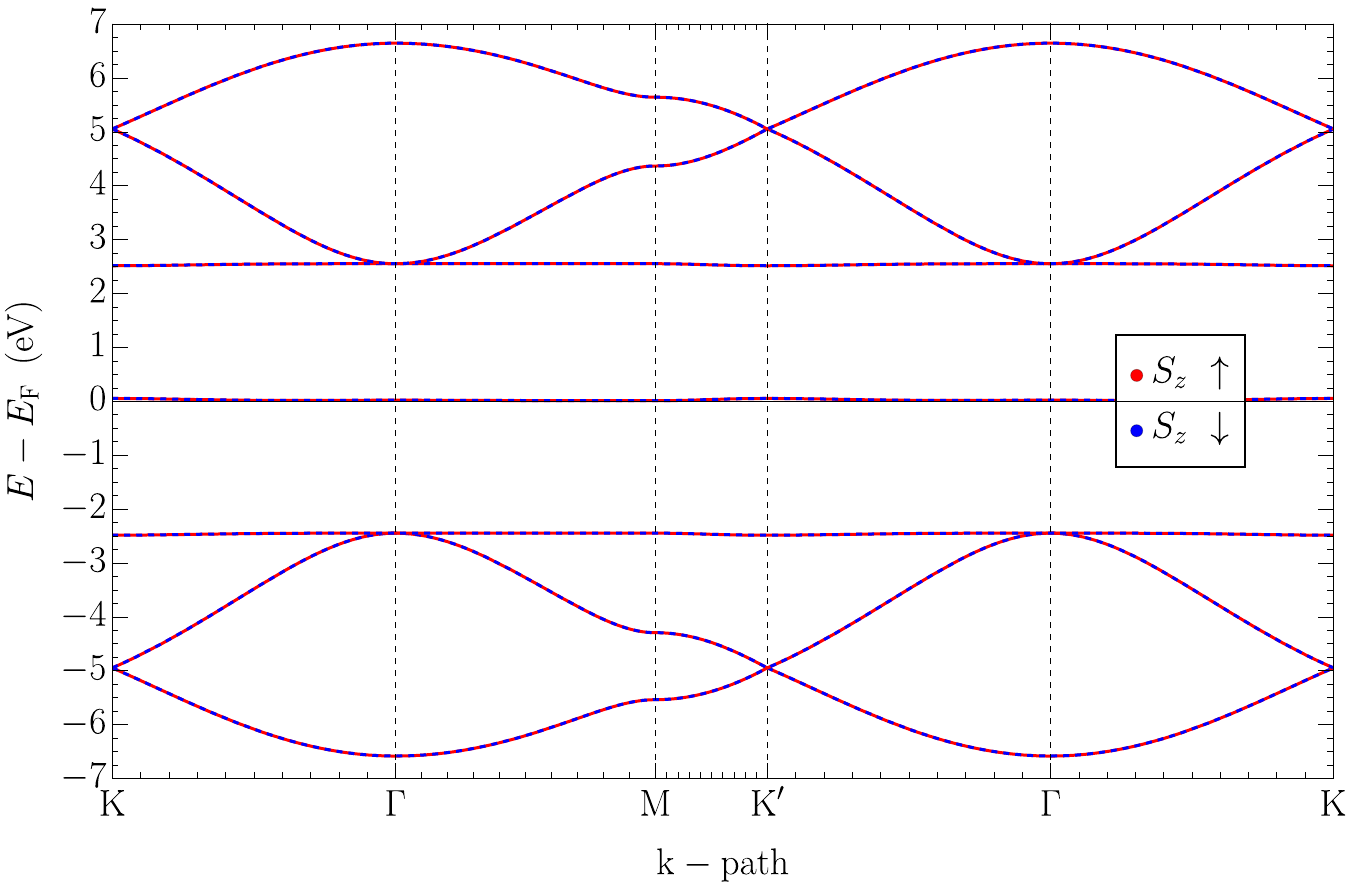}\,\includegraphics[scale=0.39]{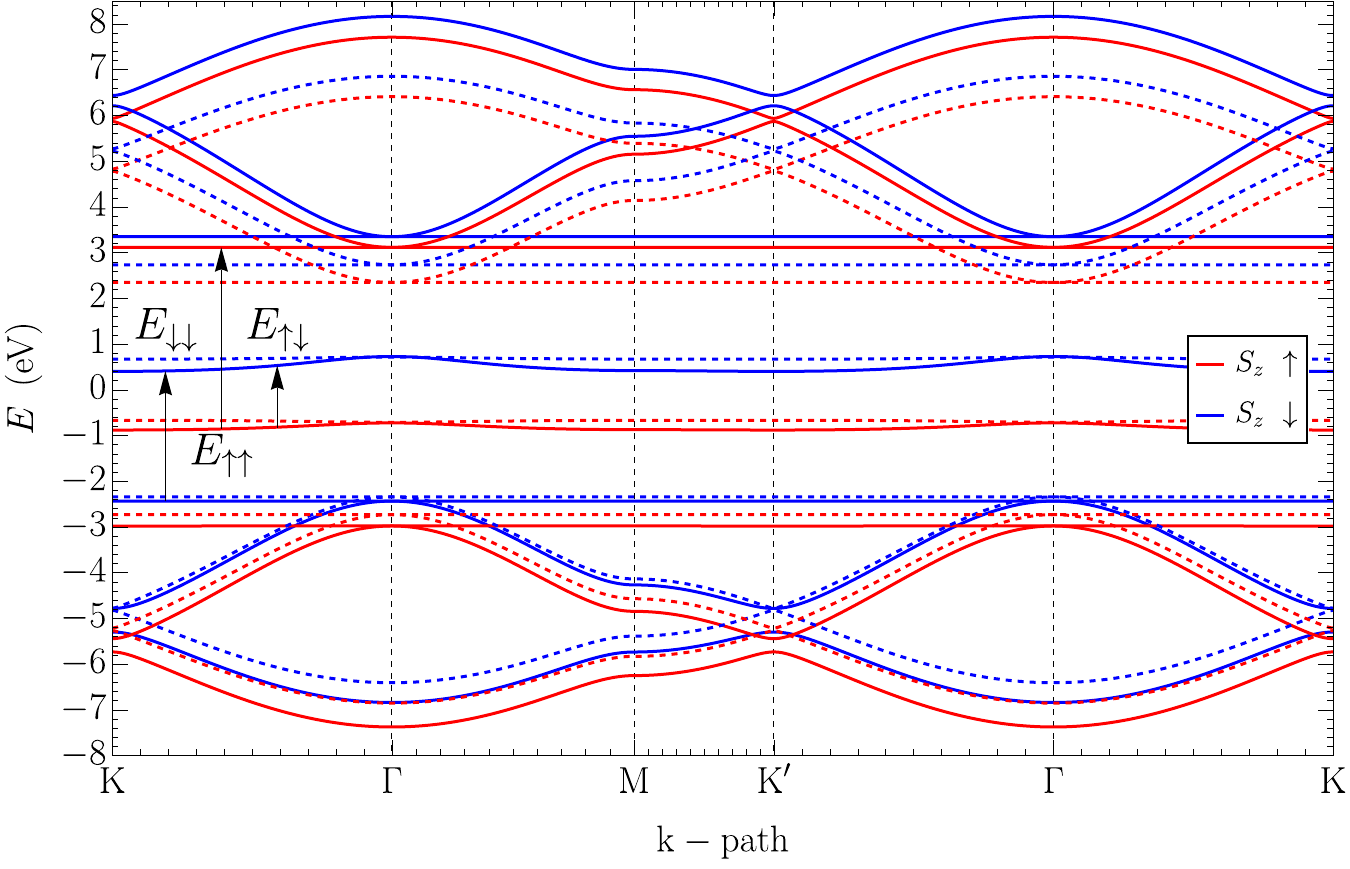}
	\vspace{-0.5cm}\centering\caption{Single-particle non-interacting (left panel) and unrestricted Hartree-Fock quasi-particle (right panel) band structure for ferromagnetic hydrogenated graphene system with $8$ Carbon atoms per unit cell. In the right hand side plot, solid lines correspond to an UHF calculation with $U=4\,\mathrm{eV}$ and $J=1.25\,\mathrm{eV}$, while dashed lines have $J=0\,\mathrm{eV}$. Quasi-particle gaps between the three distinct band pairings relevant are highlighted by the transitions in question.\label{fig:bands}}
\end{figure*}

{We first consider ferromagnetic  graphene \cite{doi:10.1126/science.aad8038,vina-bausa_building_2025} as a scenario where both excitons and magnons can be computed. 
Ferromagnetism in graphene appears when hydrogen atoms are adsorbed on the same sublattice\cite{PhysRevB.81.165409,doi:10.1126/science.aad8038,vina-bausa_building_2025} (vacancies also do the trick \cite{Palacios.PhysRevB.77.195428}, although lattice reconstructions may diminish the effect\cite{palacios_critical_2012}).
The unit cell we consider is obtained by doubling the lattice vectors of the standard graphene lattice, as plotted in Fig.~\ref{fig:lattice} and drastically increasing the on-site energy of the leftmost site in the figure to mimic the effect of the vacancy or hydrogen atom bonded to the carbon atom below. The non-interacting band structure for $t=1$ eV is  then computed, giving the expected flat band at the Fermi level\cite{Palacios.PhysRevB.77.195428}, as shown in Fig.~\ref{fig:bands}(a).}

\begin{table*}[htbp]
\centering
\footnotesize
\renewcommand{\arraystretch}{1.3}
\begin{tabular}{|c|c|c|c|c|}
\hline

\multirow{4}{*}{
    \makecell[c]{\textbf{Interaction term /}\\\textbf{bound states}}
}
&
\multicolumn{3}{c|}{\textbf{Contribution of}}
&
\multirow{4}{*}{
    \makecell[c]{\textbf{General}\\\textbf{Hamiltonian}}
}
\\
\cline{2-4}

&
\makecell[c]{\textbf{On-site}\\\textbf{interaction}}
&
\multicolumn{2}{c|}{
    \makecell[c]{\textbf{Non-local}\\\textbf{interaction $(i\neq j)$}}
}
&
\\
\cline{2-4}

&
${n}_{i\sigma}{n}_{i\bar{\sigma}}$
&
${n}_{i\sigma}{n}_{j\bar{\sigma}}$
&
\textbf{${n}_{i\sigma}{n}_{j\sigma}$}
&
\\
\hline
Hartree term in $\Sigma$ & $\neq 0$ & $\neq 0$ & $\neq 0$ & $\neq 0$ \\ 
\hline
Fock term in $\Sigma$ & $0$ & $0$ & $\neq 0$ & $\neq 0$ \\ 
\hline
$D$ for spin-conserving excitations  & $0$ & $0$ & $\neq 0$ & $\neq 0$\\
\hline
$X$ for spin-conserving excitations & $\neq 0$ & $\neq 0$ & $\neq 0$ & $\neq 0$\\
\hline
$\tilde{D}$ for spin-conserving excitations  & $0$ & $0$ & $\neq 0$ & $\neq 0$\\
\hline
$\tilde{X}$ for spin-conserving excitations & $\neq 0$ & $\neq 0$ & $\neq 0$ & $\neq 0$\\
\hline		
$D$ for spin-flip excitations  & $\neq 0$ & $\neq 0$ & $0$ & $\neq 0$\\
\hline
$X$ for spin-flip excitations  & $0$ & $0$ & $0$ & $0$\\ 
\hline  
$\tilde{D}$ for spin-flip excitations  & $\neq 0$ & $\neq 0$ & $0$ & $\neq 0$\\
\hline
$\tilde{X}$ for spin-flip excitations & $0$ & $0$ & $0$ & $0$\\
\hline
Exciton & No & \multicolumn{2}{|c|}{Yes} & Yes\\
\hline
Magnon & Yes & \multicolumn{2}{|c|}{No} & Yes\\
\hline
\end{tabular}
\caption{Contribution of on-site and non-local interactions to different terms of the BSE/TDA and BSE/RPA matrices and to the appearance of bound states.}
\label{Table:SummaryTable}
\end{table*}


{\subsection{Preliminary analysis}}

{Table~\ref{Table:SummaryTable} shows a complete dissection of how the different interaction terms in the Hamiltonian determine which terms in the self-energy and in the kernel are present or vanish. The first thing we learn is that the Fock term in the self-energy $\Sigma^{\rm X}$ vanishes if only local interactions are considered. This translates, in turn, into the vanishing of the direct term in the kernel either at the TDA or the RPA levels. {As this term is} responsible for the binding of the exciton, one concludes that excitons cannot form for the standard Hubbard model (something similar happens in the case of distinguishable particles for any interaction model, as previously discussed). The exchange term is non-zero, but does not contribute to the binding.
}
Remarkably, when considering spin-flip excitations, the finite and the vanishing terms swap roles, leading to the binding of magnons with the standard Hubbard model and their disappearance when considering only non-local interactions. 

For the calculations, we will generally consider all the {$14$} bands of the system in question {(the vacancy-related  bands are decoupled from the low energy spectrum)}. Sometimes, for illustration purposes, we will take only the $4$ bands closest to the Fermi level as to allow for easier visualization of both magnon and exciton states. This corresponds to two bands of opposite spins full and two bands of opposite spin empty (see right panel of Fig.~\ref{fig:bands}). In this plot, these three distinct transitions are highlighted, together with their respective UHF bandgaps given by
\begin{align}
    E_{\uparrow\downarrow}^{\mathrm{U}} &= 1.28\,\mathrm{eV}, & E_{\uparrow\uparrow}^{\mathrm{U}} &= 3.84\,\mathrm{eV}, & E_{\downarrow\downarrow}^{\mathrm{U}} &= 2.84\,\mathrm{eV}.\label{eq:gap-UHF}
\end{align}
For this plot, we have used $U=4$ eV and $J= 1.25$ eV.

\begin{figure}
	\centering\includegraphics[scale=0.385]{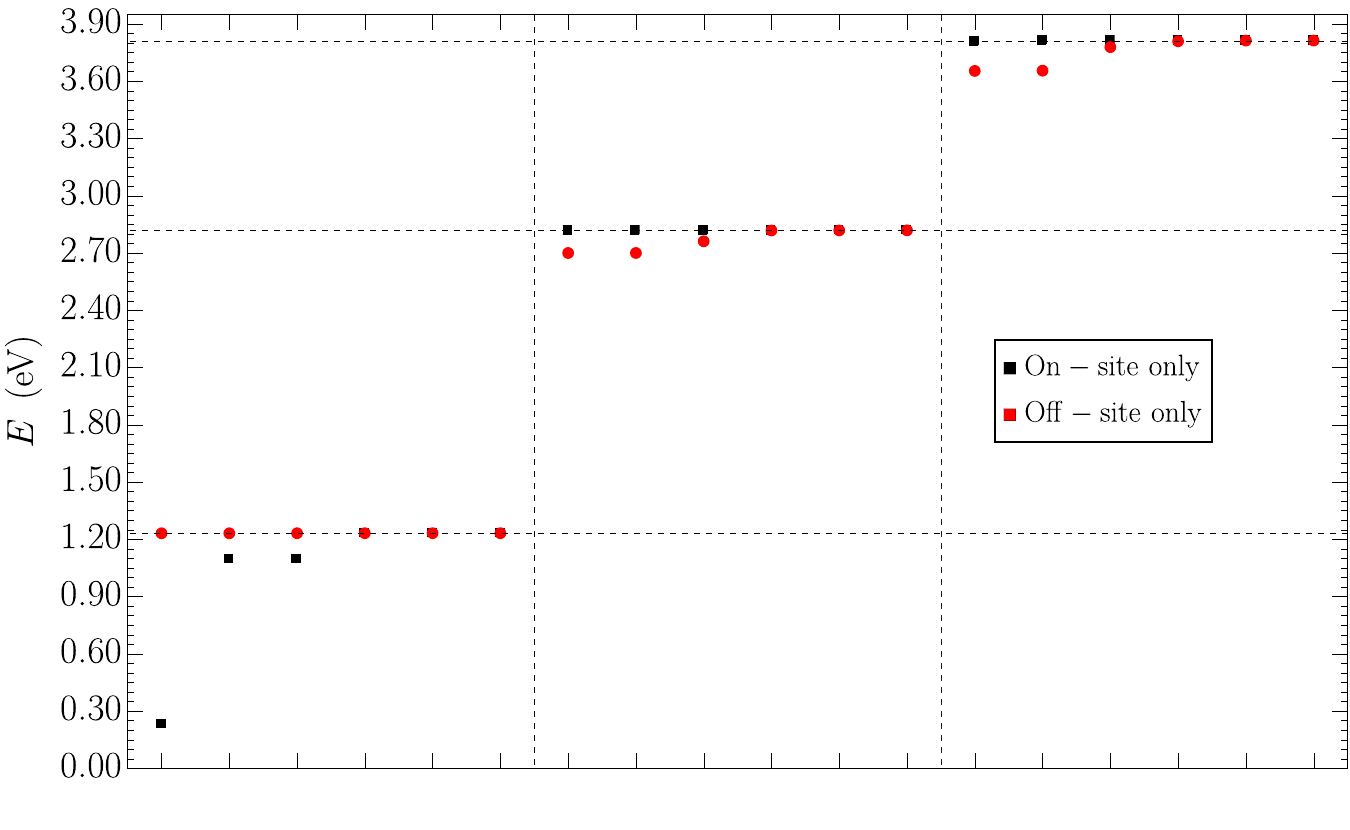}
	\vspace{-0.5cm}\centering\caption{Effects of including only on-site versus off-site on the excitation energies in the BSE/TDA formalism. First six eigenvalues included in each spin sector, separated by the vertical dashed lines as $\uparrow\downarrow$, $\downarrow\downarrow$, and $\uparrow\uparrow$. Horizontal dashed lines correspond to the bandgap for each of these sectors. 
    \label{fig:UvsJ}}
\end{figure}

{We have verified our previous observations by performing calculations setting $U$ or $J$ to zero independently {on top of the same ground state discussed above}.
Setting $U=0$ and keeping a finite value of $J$ (in this case, $1.25\,\mathrm{eV}$), we observe that magnons are completely lost and only continuum spin-flip excitations (\emph{i.e.}, Stoner excitations \cite{10.1098/rspa.1938.0066,nag_correlation_2022}) remain in that energy range, while excitons remain largely unaffected. 
On the contrary, setting $J=0$ and keeping a finite value for $U=4\,\mathrm{eV}$, excitons are lost and only continuum spin-conserving excitations remain, while magnons remain untouched}. This is all shown in Fig.~\ref{fig:UvsJ}.

\subsection{BSE/TDA with unrestricted Hartree-Fock ground state\label{sec:uhf}}

Although  UHF is a standard MF approach for magnetic systems, it  suffers from spin contamination \cite{andrews_spin_1991,nandi_effect_1996,young_spin_2001,rodriguez-guzman_symmetry-projected_2012,rodriguez-guzman_multireference_2013} as $\ket{\Phi_0}$ is not an eigenstate of $ S^2$ (it remains an eigenstate of $ S_z$). Its spin contamination $\Delta\langle S^2\rangle$ is given by Eq.~\ref{eq:ynorm}, as we have shown above.

\begin{figure}
	\centering\includegraphics[scale=0.4]{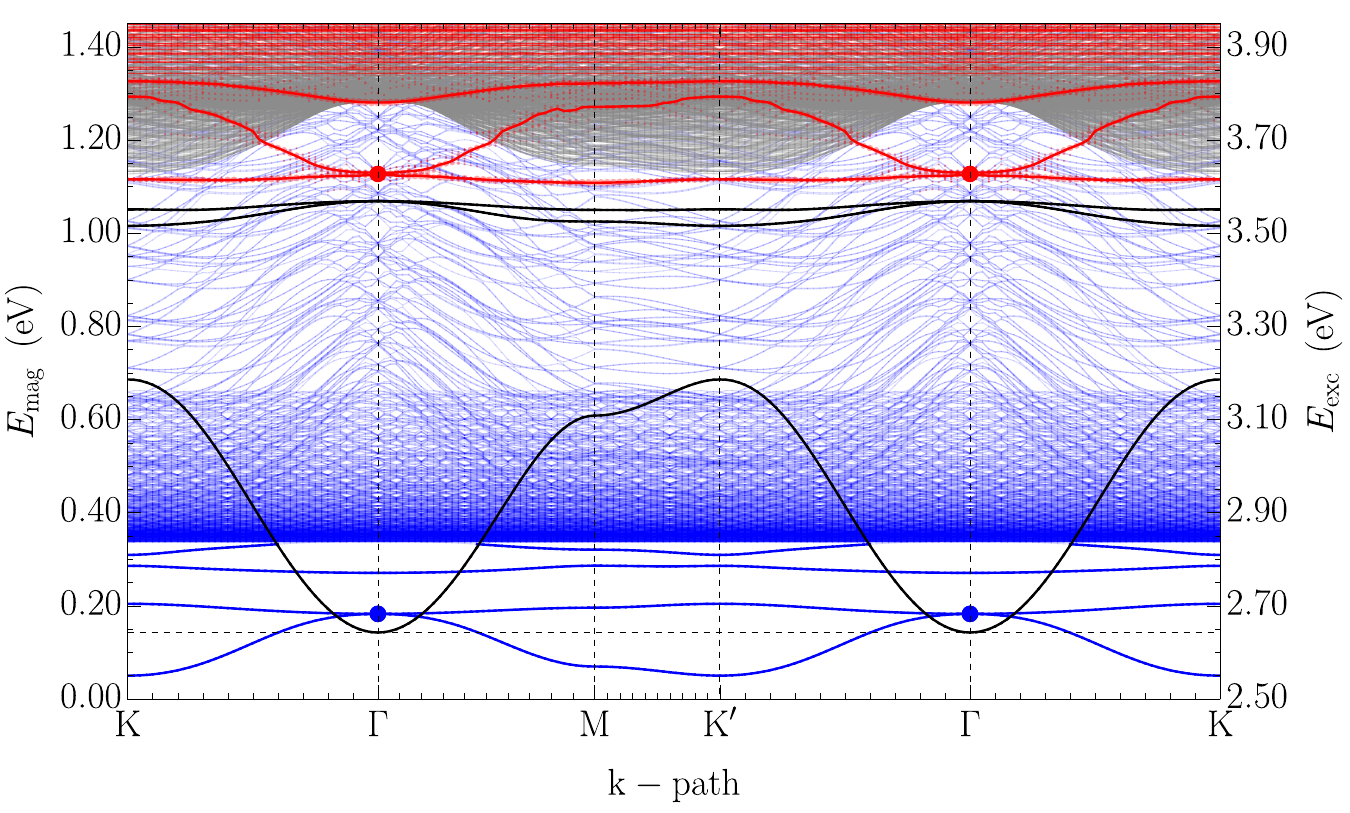}
	\vspace{-0.3cm}\centering\caption{
    Magnon (black lines) and exciton (blue and red lines) dispersion for COM momentum $\bm{Q}$ along the $\boldsymbol{\mathrm{K}}$-$\boldsymbol{\Gamma}$-$\boldsymbol{\mathrm{M}}$-$\boldsymbol{\mathrm{K}}'$-$\boldsymbol{\Gamma}$-$\boldsymbol{\mathrm{K}}$ path in the first Brillouin zone with BSE/TDA. Black lines highlight the magnon Goldstone mode and the optical magnons, grey lines are the Stoner continuum. Solid dots at $\boldsymbol{\Gamma}$ in the exciton lines denotes an optically bright state. Excitons depicted in blue/red correspond to transitions associated with the band pairings denoted as $E_{\downarrow\downarrow}/E_{\uparrow\uparrow}$ in Fig.~(\ref{fig:bands}), respectively.
    \label{fig:magnon-exciton}}
\end{figure}

For now, let us temporarily ignore this spin contamination and proceed with the BSE/TDA computations knowing that a gap in the Goldstone mode will be present. One obtains the excitation dispersion as shown in Fig.~\ref{fig:magnon-exciton}. Besides the (gapped) Goldstone mode (black line), one clearly observes two other bound magnons degenerate at $\boldsymbol{\Gamma}$ (also in black, almost flat dispersion), usually called "optical" modes \cite{PhysRevLett.54.227,CookeBlackmanMorgan1985,PhysRevB.58.293,CookeLynnDavis1980,gbpw-zh1v}, followed by the Stoner continuum (grey lines). Looking further up in the energy series, we reach spin-conserving excitations in the form of excitons, first for spin-down transitions (blue lines) and then for spin-up transitions (red lines).  The excitonic continuum and transitions corresponding to different band pairings (for instance, the spin-down band which touches one of the highlighted bands at $\boldsymbol{\Gamma}$, recall the right panel of Fig.~\ref{fig:bands}) are also clearly visible, filling the space between the two excitonic sectors. 

Regarding the optically bright/dark nature of each state, magnons are optically dark as expected while excitons can be bright or dark depending on their dipole matrix elements. This is highlighted in Fig.~\ref{fig:magnon-exciton}, where optically bright excitons are identified with a solid dot. Furthermore, in Fig.~\ref{fig:optical-response} we plot the linear optical conductivity of this system, clearly observing both the spin selectivity of optical transitions and the highly localized low energy peaks,  characteristic of an almost flat band structure.

\subsection{BSE/RPA with UHF ground state\label{sec:rpa_uhf}}

\begin{figure}
	\centering\includegraphics[scale=0.385]{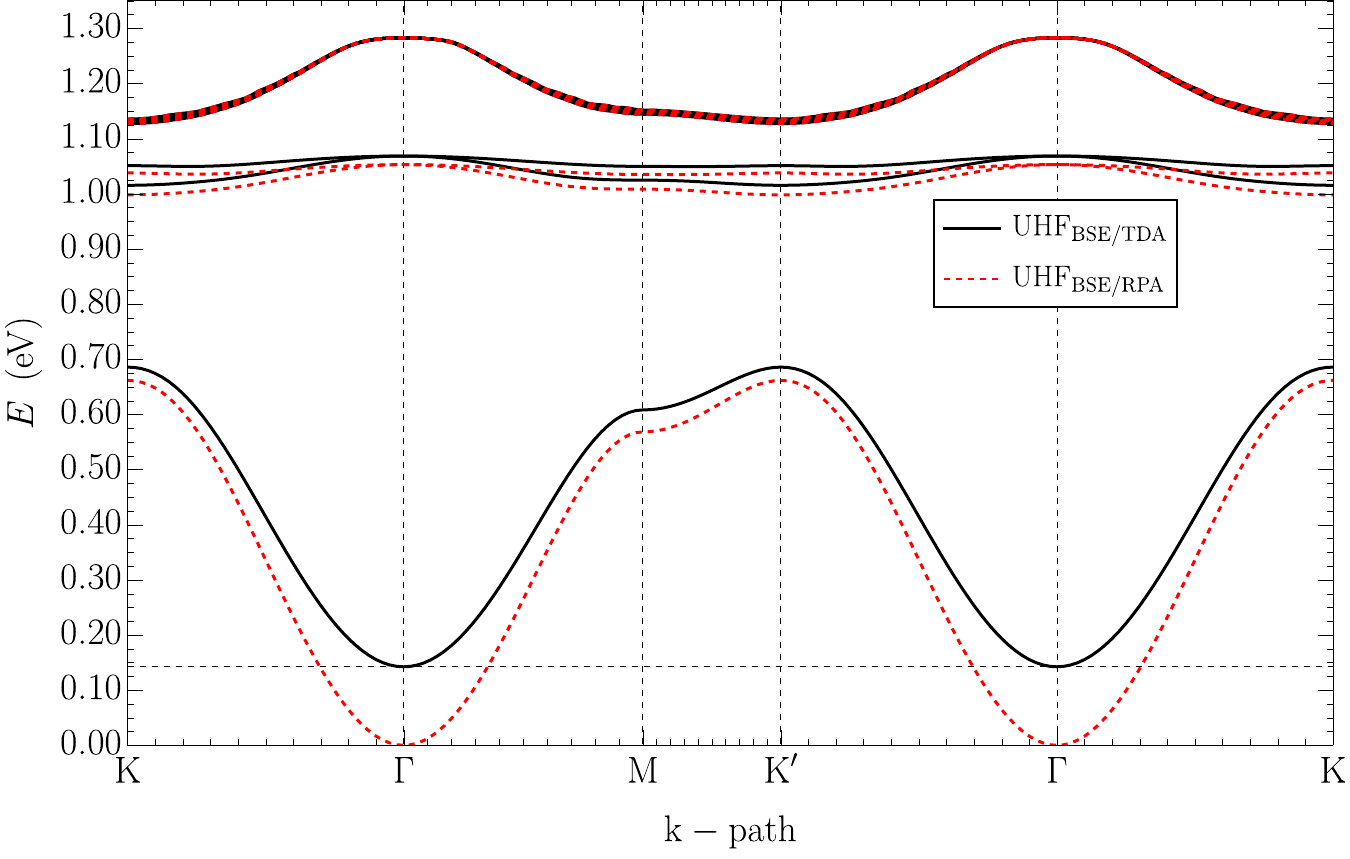}
	\vspace{-0.5cm}\centering\caption{Magnon dispersion from UHF computed via BSE/TDA (black lines) or BSE/RPA (red dashed lines) for COM momentum $\bm{Q}$ along the $\boldsymbol{\mathrm{K}}$-$\boldsymbol{\Gamma}$-$\boldsymbol{\mathrm{M}}$-$\boldsymbol{\mathrm{K}}'$-$\boldsymbol{\Gamma}$-$\boldsymbol{\mathrm{K}}$ path in the first Brillouin zone.
    \label{fig:UHFvsBTDA}}
\end{figure}

Recalling Secs. \ref{sec:btda} and \ref{sec:goldstone_btda}, we will now show that BSE/RPA {fully closes} the spin-contamination gap which appears in the Goldstone mode. Including the anti-resonant excitations and their coupling, the Goldstone mode becomes exactly gapless, as can be seen in Fig.~\ref{fig:UHFvsBTDA}. The spin contamination of the UHF reference itself is unchanged, but it no longer gaps the Goldstone mode, as discussed in Eq.~\ref{eq:ynorm}.
Furthermore, while no {change} is observed in the Stoner continuum, the optical magnon modes also get shifted down in energy in similar fashion to the Goldstone mode. At small $\bm Q$ the Goldstone mode is quadratic with dispersion
\begin{equation}
	\hbar\omega(\bm Q)\simeq 2.307\,\mathrm{eV}\text{\AA}^2 \,|\bm Q|^2,
\end{equation}
as expected from {the magnetization-based classification from} Eq.~\ref{eq:znorm}, against a gap at $\boldsymbol{\Gamma}$ of $0.14\,\mathrm{eV}$ for BSE/TDA. Additionally, it is clear that the energy difference is not a pure scissor cut, being instead dependent on both $\bm{Q}$ and on which state one is discussing. 

Regarding the exciton states, we have omit their dispersion to allow for a clearer observation of the exactly-zero nature of the Goldstone magnon. However, we still compare the (optically active) excitons in both BSE forms via the linear optical conductivity, as presented in Fig.~\ref{fig:optical-response}. Consistent with previous work in the literature \cite{sander_beyond_2015}, the effect of including anti-resonant excitations and their coupling to resonant excitations in the linear optical conductivity is negligible. Regarding dark excitons with $\Delta S_z =1$, as mentioned previously, these are outside of the range plotted due to the $\approx 5\,\mathrm{eV}$ energy difference between the bands involved in such transitions. For another comparison between the two forms of the BSE, we showcase the excitations' $\bm{k}$-resolved state weights (analogous to the wave function for BSE/TDA) in Appendix~\ref{sec:wavefunctions}.

\section{Antiferromagnetic Graphene}

\begin{figure}
    \centering\includegraphics[scale=0.39]{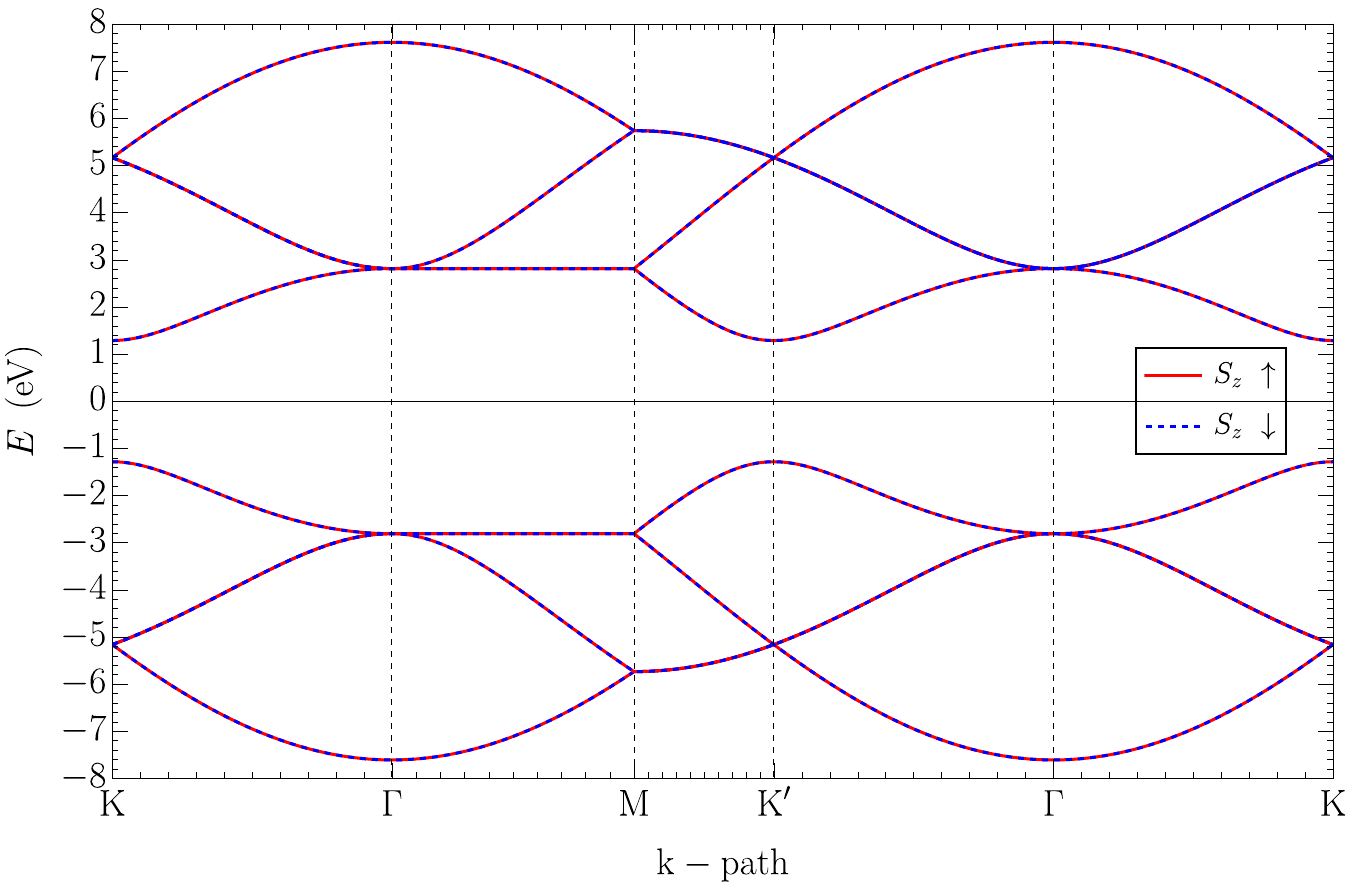}
	\vspace{-0.5cm}\centering\caption{Unrestricted Hartree-Fock quasi-particle band structure for an antiferromagnetic graphene system with $8$ Carbon atoms per unit cell.
    \label{fig:bands_AFM}}
\end{figure}

Having shown that the Goldstone mode can be correctly computed in a ferromagnet, we will now consider the case of antiferromagnetic graphene. Considering the same unit cell as before, namely the one obtained by doubling the lattice vectors of the standard graphene lattice, as shown in Fig.~\ref{fig:lattice}, as well as the same model interaction as presented in Eq.~\ref{eq:gen_hubbard}, we now perform the UHF calculation for an antiferromagnetic ground state. Even for pristine graphene, this ground state appears when the on-site interaction strength reaches $U=7\,\mathrm{eV}$, with band structure shown in Fig.~\ref{fig:bands_AFM}. {For this specific case, we keep $J=0$.}

\begin{figure}
	\centering\includegraphics[scale=0.385]{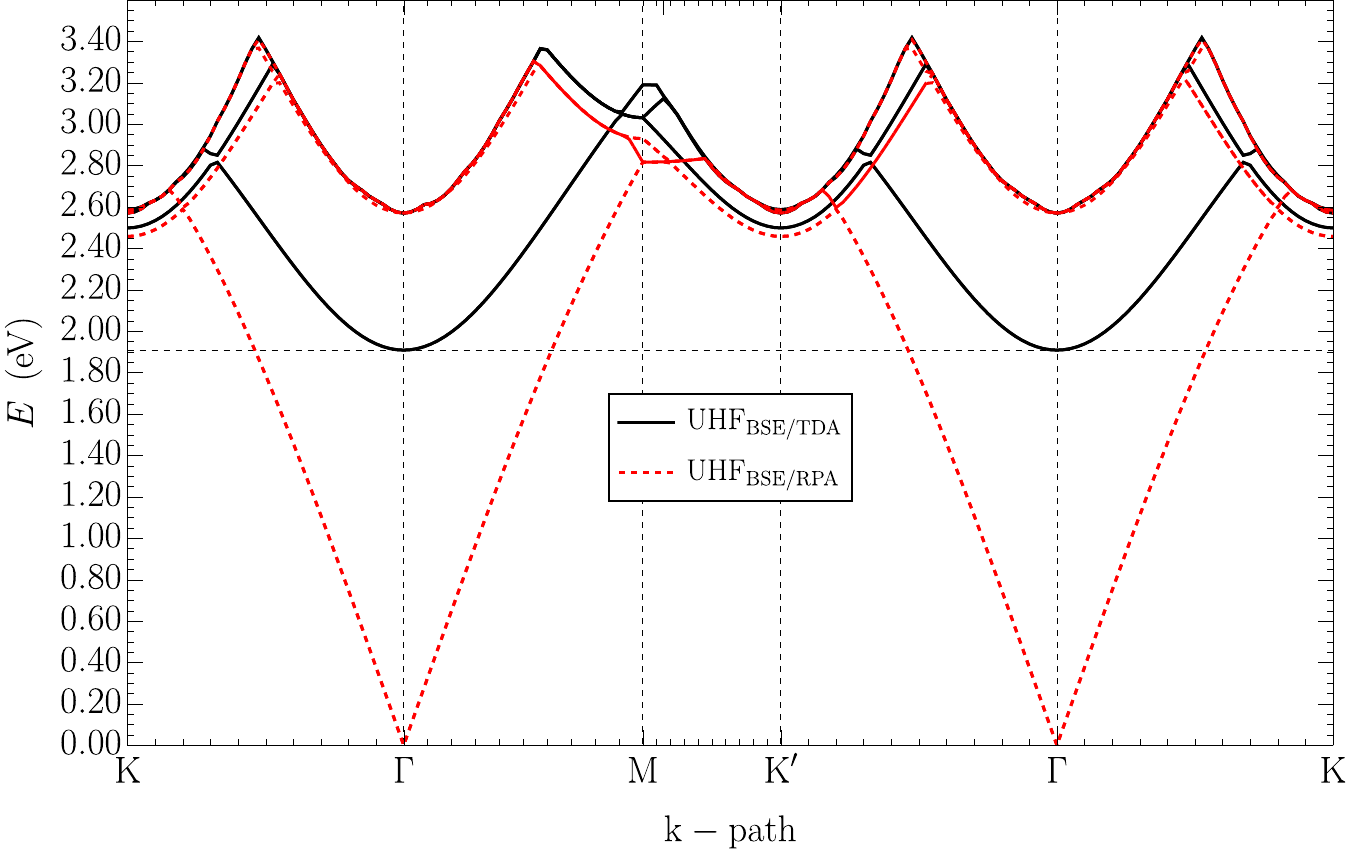}
	\vspace{-0.5cm}\centering\caption{Magnon dispersion from UHF computed via BSE/TDA (black lines) or BSE/RPA (red dashed lines) for COM momentum $\bm{Q}$ along the $\boldsymbol{\mathrm{K}}$-$\boldsymbol{\Gamma}$-$\boldsymbol{\mathrm{M}}$-$\boldsymbol{\mathrm{K}}'$-$\boldsymbol{\Gamma}$-$\boldsymbol{\mathrm{K}}$ path in the first Brillouin zone. 
    \label{fig:UHFvsBTDA_AFM}}
\end{figure}

With $\langle S_z\rangle=0$, Eq.~\ref{eq:znorm} gives two zero-norm zero modes at $\boldsymbol\Gamma$ and the BSE/RPA Goldstone magnon is linear and twofold degenerate with dispersion 
 \begin{equation}
 	\hbar\omega(\bm Q)\simeq 4.42\,\,\mathrm{eV}\text{\AA}\,|\bm Q|,
 \end{equation} 
{close to $\boldsymbol\Gamma$} against a $1.91\,\mathrm{eV}$ gap in the BSE/TDA as shown in Fig.~\ref{fig:UHFvsBTDA_AFM}.
While in the ferromagnetic case the difference in magnon dispersion between BSE/TDA and BSE/RPA might seem, at first glance, like a simple downward shift in energy, here the difference is much more drastic. Besides the much larger energy difference between the two at $\boldsymbol{\Gamma}$, the curves are also qualitatively distinct, with BSE/TDA failing to capture the linear dispersion of the Goldstone magnons and presenting a quadratic-like dispersion similar to the ferromagnetic case. BSE/RPA, on the contrary, accurately captures the Goldstone magnons' linear dispersion and the two exact zero modes at $\boldsymbol{\Gamma}$.

\section{Conclusions}

In this paper, we have shown that magnons and excitons can be computed on the same footing under the Bethe-Salpeter formalism while respecting the gapless Goldstone mode for magnons. This result rests on working beyond the Tamm-Dancoff approximation through the evaluation of the full Bethe-Salpeter equation with the same interaction in the kernel as that which was used in the computation of the starting mean-field ground state. This follows directly from the equation-of-motion derivation of the Bethe-Salpeter equation and the Goldstone theorem, and we have shown that keeping the anti-resonant block and its coupling to the resonant block of the Bethe-Salpeter matrix is fundamental if one wishes to preserve its zero modes. 

Following this procedure, we have shown that the vestigial Goldstone gap, present throughout the current literature and usually closed by a manual scissor cut in ferromagnets, closes fully and the exact zero-modes are recovered. 
Additionally, we have also shown that the correction to the magnon dispersion is not purely a scissor cut, being strongly dependent on which state one is analysing, on the centre-of-mass momentum $\bm{Q}$, and on the magnetic phase of the material. This is especially noticeable when one discusses antiferromagnets, where the linear Goldstone mode is completely lost in the Tamm-Dancoff approximation, instead becoming quadratic and presenting a sizeable gap. 

Moving on to excitons, we have verified that the Tamm-Dancoff approximation is very accurate for these states, matching the observations of previous works that the changes in the excitonic optical response are minimal. Furthermore, the connection between the existence of both excitons and magnons and the presence of on-site and non-local interactions, as well as their respective roles in the Bethe-Salpeter kernel, has been discussed and verified.

\begin{figure*}
    \centering\includegraphics[scale=0.745]{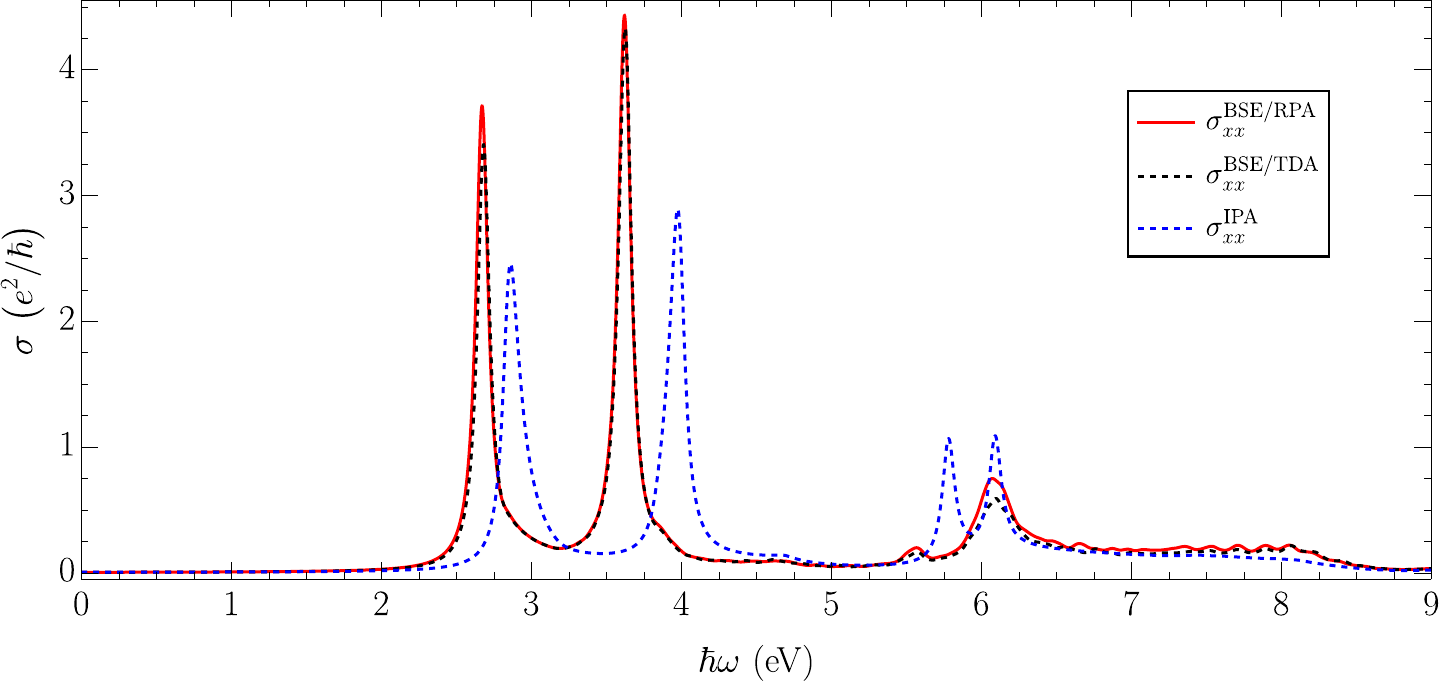}
    \vspace{-0.7cm}\centering\caption{Excitonic (red and black) and IPA (blue) contributions to the diagonal tensor elements of the linear optical conductivity for ferromagnetic graphene.
    As expected, only spin-conserving transitions are optically active for both contributions. Red and black lines correspond to excitons computed via BSE/RPA and BSE/TDA, respectively. 
    \label{fig:optical-response}}
\end{figure*}

\section*{Acknowledgments}
We acknowledge financial support from MICIU/AEI/10.13039/501100011033 through Grants PID2022-141712NB-C21, PCI2026-177428-1,  PID2025-174732NB-I00, and CEX2023-001316-M within the María de Maeztu Program for Units of Excellence in R\&D. We acknowledge the Comunidad de Madrid through the project ``Disruptive 2D materials'' (MAD2D-CM-UAM7) within the Recovery, Transformation and Resilience Plan. We acknowledge the European Union through the M.ERA-NET Joint Call 2025 (ALTMAG project) and the NextGenerationEU programme. We also acknowledge computational resources from CCC at UAM and the Spanish RES through projects FI-2025-1-12 and FI-2026-2-0015.

\section*{Author Contributions}
Simran Chourasia and Maur{\'i}cio F. C. Martins Quintela contributed equally to this work. J. J. Palacios conceived and supervised the work. All authors contributed equally to writing the manuscript.

\begin{appendices}

\section{Distinguishable electrons\label{sec:distinguishable}}

In this appendix, we outline the derivation of the BSE/TDA considering distinguishable electrons. Recalling the definitions of the main text
\begin{align}
	{{H}} & = T +  {V}, \\
	T & = \sum_{l_1, l_2} T_{l_1 l_2} {c}_{l_1}^\dagger  {c}_{l_2} = \sum_{\alpha} \varepsilon_{\alpha} {a}_{\alpha}^\dagger {a}_{\alpha}, \\
	{V} & = \frac{1}{2}\sum_{l_1, l_2, l_3, l_4} V_{l_1 l_2 l_3 l_4} {c}_{l_1}^\dagger  {c}_{l_2}^\dagger {c}_{l_4} {c}_{l_3} ,
\end{align}
the ground state is assumed as $\ket{\Phi_0} = \prod_{\varepsilon_v \leq E_F} {a}_{v}^\dagger \ket{0}$. Here, subscript $v$ and $c$ represent valence and conduction band state, respectively.
In the e-h excitation basis, $\ket{v, c} = {a}_{c}^\dagger {a}_{v} \ket{\Phi_0}$, the Hamiltonian becomes
\begin{align} \label{TDA_HF}
    & \bra{v, c} {{H}} \ket{v',c'}= \delta_{cc'} \delta_{vv'} \left[ \bra{\Phi_0}{{H}}\ket{\Phi_0} + \varepsilon_c - \varepsilon_v \right]  \nonumber \\
    & - \delta_{cc'} \sum_{\varepsilon_\alpha \leq E_F} \left[ \bra{\varepsilon_{v'}, \varepsilon_\alpha} {V} \ket{\varepsilon_v, \varepsilon_\alpha} - \bra{\varepsilon_\alpha, \varepsilon_{v'}} {V} \ket{\varepsilon_v, \varepsilon_\alpha} \right] \nonumber \\
	&+ \delta_{vv'} \sum_{\varepsilon_\alpha \leq E_F} \left[ \bra{\varepsilon_c, \varepsilon_\alpha} {V} \ket{\varepsilon_{c'}, \varepsilon_\alpha} - \bra{\varepsilon_c, \varepsilon_\alpha} {V} \ket{\varepsilon_\alpha, \varepsilon_{c'}} \right] \nonumber \\	
	& - \bra{\varepsilon_c, \varepsilon_{v'}} {V} \ket{\varepsilon_{c'}, \varepsilon_v} + \bra{\varepsilon_c, \varepsilon_{v'}} {V} \ket{\varepsilon_v, \varepsilon_{c'}}. 
\end{align}

A system with non-identical interacting electrons will only have Hartree interaction. The Fock interaction term appears because all the electrons are identical. Since the Fock term carries all the information of many-body correlations, it is important to know which terms of the kernel arise from it. Looking at TDA matrix elements for non-identical electrons will help us identify these terms. To study non-identical electrons, let us define the creation operator ${c}^\dagger_{l}(x)$ which creates the electron with label "$x$" in the electronic state $l$, and ${c}_{l}(x)$ is the corresponding annihilation operator. Now, the Hamiltonian for non-identical (NI) electrons with two-body interaction has the following non-interacting and interaction terms
\begin{align}
    {{H}}^\text{NI} & = T^\text{NI} +  {V}^\text{NI}, \\
	T^\text{NI} & = \sum_{x_1} \sum_{l_1, l_2} T_{l_1 l_2} {c}_{l_1}^\dagger (x_1) {c}_{l_2}(x_1) \nonumber\\
    & = \sum_{x_1} \sum_{\alpha} \varepsilon_{\alpha} {a}_{\alpha}^\dagger (x_1) {a}_{\alpha}(x_1), \\
	{V}^\text{NI} & = \frac{1}{2}\sum_{x_1, x_2} \sum_{l_1, l_2, l_3, l_4} \big[ (1- \delta_{x_1 x_2}) V_{l_1 l_2 l_3 l_4}(x_1, x_2) \nonumber\\
    & {c}_{l_1}^\dagger (x_1) {c}_{l_2}^\dagger (x_2) {c}_{l_4}(x_2) {c}_{l_3}(x_1) \big].
\end{align}
The N-electron ground takes the form
\begin{align}
	\ket{\Phi_0^\text{NI}} = & \ket{\varepsilon_{v_1}(p_1), \varepsilon_{v_2}(p_2), \dots, \varepsilon_{v_N}(p_N)} \nonumber\\
    = & {a}^\dagger_{v_1}(p_1) {a}^\dagger_{v_2}(p_2) \dots {a}^\dagger_{v_N}(p_N) \ket{0},
\end{align}
and the Hamiltonian in the e-h excitation basis, $\ket{v_i(p_i), c(\eta)} = {a}_{c}^\dagger(\eta) {a}_{v_i}(p_i) \ket{\Phi_0}$, becomes
\begin{align} \label{TDA_Hartree}
	 & \bra{v_i(p_i), c(\eta)} {{H}}^\text{NI} \ket{v_j(p_j), c'(\eta')} \nonumber \\ 
	& =\delta_{c c'} \delta_{\eta \eta'} \delta_{v_i v_j} \delta_{p_i p_j}  \Big[ \bra{\Phi_0^\text{NI}} {{H}}^\text{NI} \ket{\Phi_0^\text{NI}} + \varepsilon_{c}- \varepsilon_{v_i}\Big] \nonumber\\
	& - \delta_{c c'} \delta_{\eta \eta'} \delta_{v_i v_j} \delta_{p_i p_j}\sum_{r=1}^N \Big[ (1-\delta_{v_r v_i} \delta_{p_r p_i}) \nonumber\\ 
    & \bra{\varepsilon_{v_r}(p_r), \varepsilon_{v_i}(p_i)} {V}^\text{NI} \ket{\varepsilon_{v_r}(p_r), \varepsilon_{v_i}(p_i)} \Big] + \delta_{\eta \eta'} \delta_{v_i v_j} \delta_{p_i p_j} \nonumber\\
	& \sum_{r=1}^N (1-\delta_{p_r \eta}) \bra{\varepsilon_{v_r}(p_r), \varepsilon_c(\eta)} {V}^\text{NI} \ket{\varepsilon_{v_r}(p_r), \varepsilon_{c'}(\eta)} \nonumber \\
	& - \delta_{\eta \eta'} \delta_{v_i v_j} \delta_{p_i p_j} (1-\delta_{p_i \eta}) \nonumber \\
    &\bra{\varepsilon_{v_i}(p_i), \varepsilon_c(\eta)} {V}^\text{NI} \ket{\varepsilon_{v_i}(p_i), \varepsilon_{c'}(\eta)} \nonumber \\
	& + (1 - \delta_{v_j v_i} \delta_{p_j p_i}) \delta_{\eta' p_j} \delta_{\eta p_i} \nonumber \\ 
    &\bra{\varepsilon_{v_j}(p_j), \varepsilon_c(p_i)} {V}^\text{NI} \ket{\varepsilon_{c'}(p_j), \varepsilon_{v_i}(p_i)}.
\end{align}
To compare the above expression with Eq.~\ref{TDA_HF}, let us suppress the particle labels and replace $v_i \rightarrow v$, $v_j \rightarrow v'$, and $\sum_{r=1}^N \rightarrow \sum_{\varepsilon_{\alpha} \leq E_F}$. The resulting expression is
\begin{align}
	 & \bra{v, c} {{H}}^\text{NI} \ket{v', c'} = \delta_{c c'} \delta_{v v'} \Big[ \bra{\Phi_0^\text{NI}} {{H}}^\text{NI} \ket{\Phi_0^\text{NI}} + \varepsilon_{c}- \varepsilon_{v}\Big] \nonumber\\
	& - \delta_{c c'} \delta_{v v'} \sum_{\varepsilon_{\alpha} \leq E_F} \Big[ (1-\delta_{\alpha v}) \bra{\varepsilon_{\alpha},\varepsilon_{v}}{V}^\text{NI} \ket{\varepsilon_{\alpha}, \varepsilon_{v}} \Big] \nonumber\\
	& + \delta_{v v'} \sum_{\varepsilon_{\alpha} \leq E_F} \bra{\varepsilon_{\alpha},\varepsilon_c} {V}^\text{NI} \ket{\varepsilon_{\alpha}, \varepsilon_{c'}} \nonumber \\
	& - \delta_{v v'} \bra{\varepsilon_c, \varepsilon_{v}} {V}^\text{NI} \ket{\varepsilon_{c'}, \varepsilon_{v}} + (1 - \delta_{v' v}) \bra{\varepsilon_c, \varepsilon_{v'}} {V}^\text{NI} \ket{\varepsilon_{v}, \varepsilon_{c'}},
\end{align}
and it takes the form of Eq.~\ref{eq:TDA_NI} for a crystal.

\section{Wave functions\label{sec:wavefunctions}}	

In this appendix, we provide a visual comparison between the BSE/TDA and BSE/RPA excitation $\bm{k}$-resolved state weight in the symplectic ($\eta$) norm, which we will now designate as ``wave function'' for simplicity. From Eqs.~\ref{eq:rpa_excitation} and~\ref{eq:znorm}, it reads
\begin{align}
  \rho_{\bm{Q}}(\bm{k}) &= \psi^{\dagger}_{\bm{Q}}(\bm{k})\,\eta\,\psi_{\bm{Q}}(\bm{k})\nonumber\\
  & = \sum_{vc}\Bigl(\bigl|A^{\bm{Q}}_{cv}(\bm{k})\bigr|^{2} - \bigl|B^{\bm{Q}}_{cv}(\bm{k})\bigr|^{2}\Bigr),
  \label{eq:rho}
\end{align}
with $\psi_{\bm{Q}}(\bm{k}) = \bigl(A^{\bm{Q}}(\bm{k}),\,B^{\bm{Q}}(\bm{k})\bigr)^{T}$. This quantity is defined analogously to the BSE/TDA case \cite{uria-alvarez_efficient_2024}, but now taking advantage of the symplectic ($\eta$) normalization of the BSE/RPA eigenstates. For $B = 0$ it reduces to the BSE/TDA weight $\sum_{vc}\bigl|A^{\bm{Q}}_{cv}(\bm{k})\bigr|^{2}$, the probability of finding the electron-hole pair at $\bm{k}$.

\begin{figure*}
	\centering\includegraphics[scale=0.5]{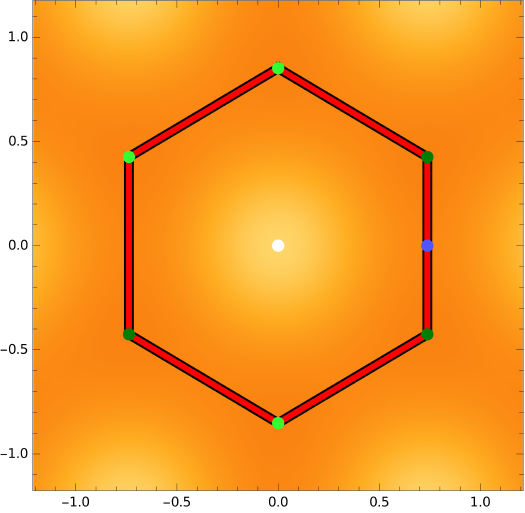}\,\includegraphics[scale=0.5]{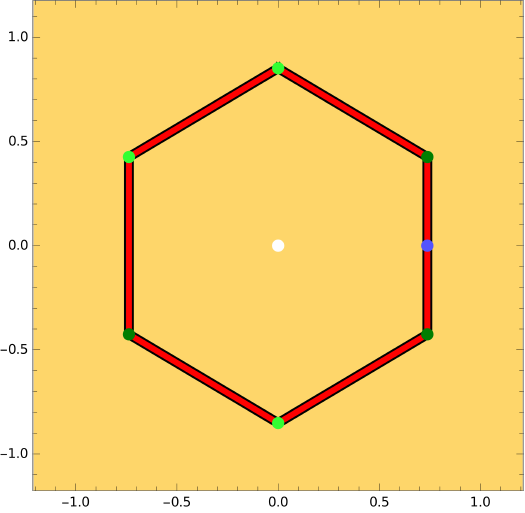}\,\includegraphics[scale=0.5]{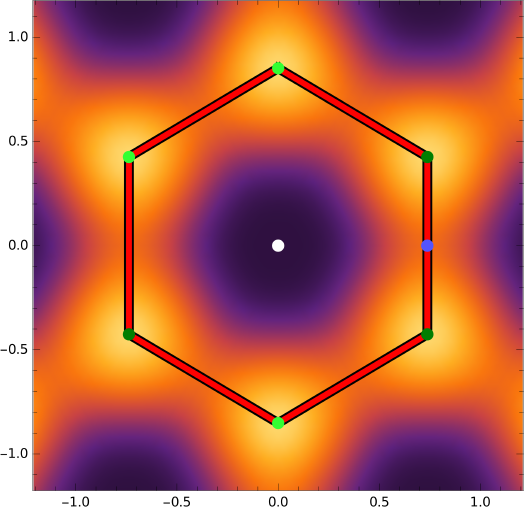}\,\includegraphics[scale=0.5]{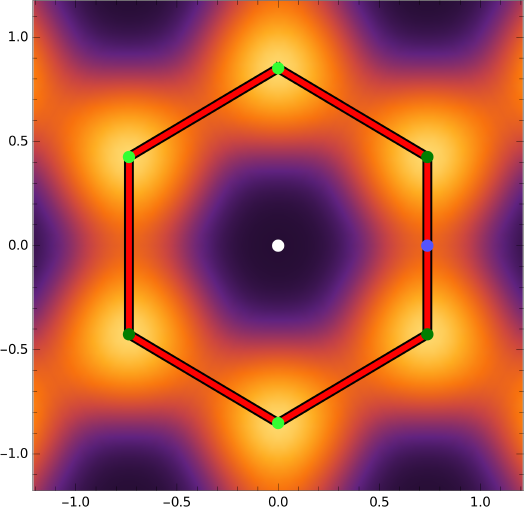}
	
	\centering\includegraphics[scale=0.5]{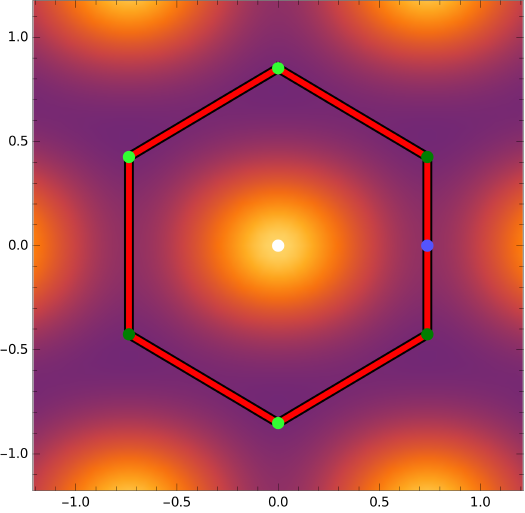}\,\includegraphics[scale=0.5]{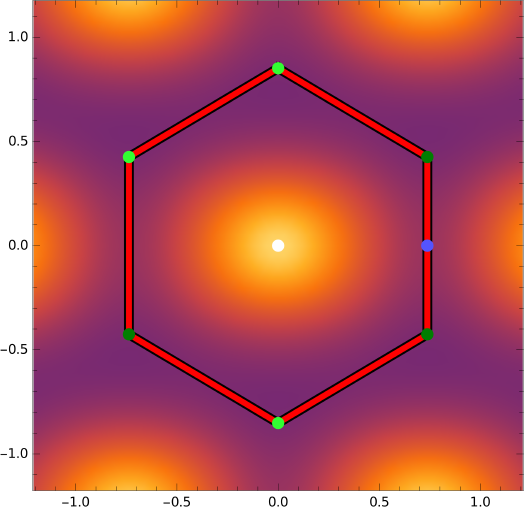}\,\includegraphics[scale=0.5]{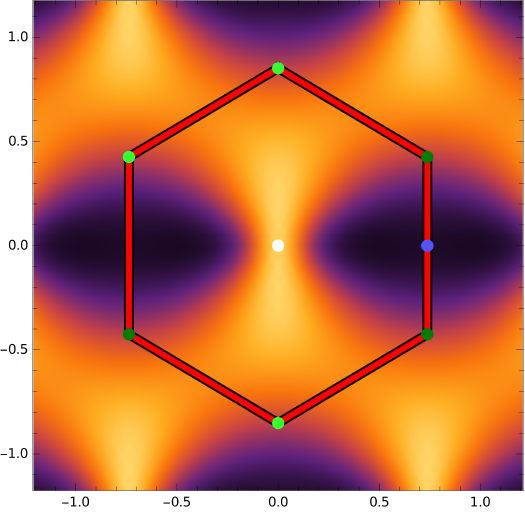}\,\includegraphics[scale=0.5]{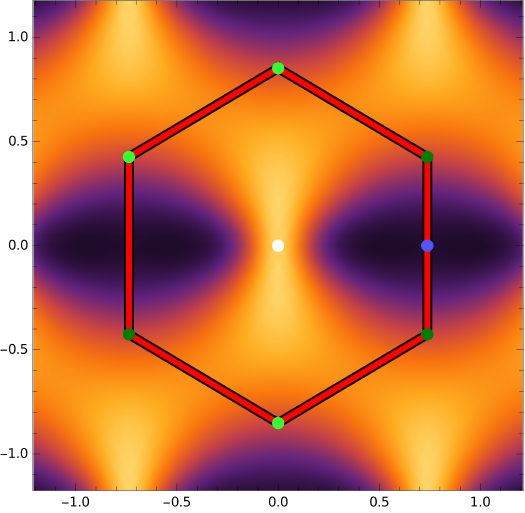}
	
	\centering\includegraphics[scale=0.5]{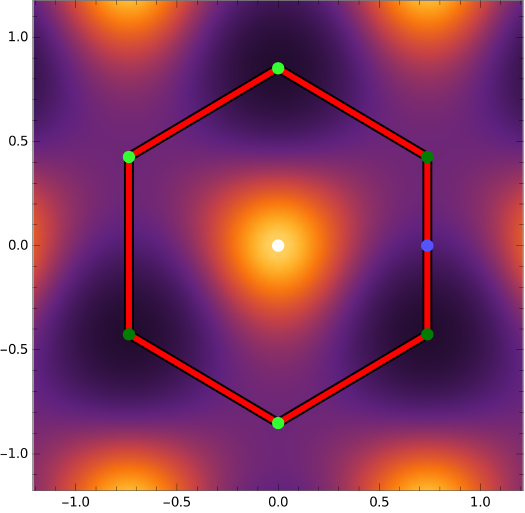}\,\includegraphics[scale=0.5]{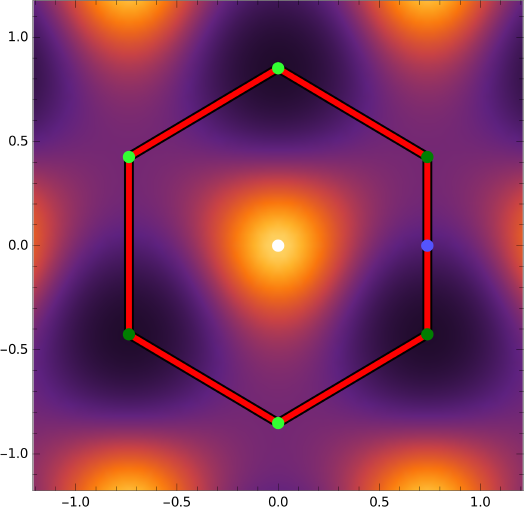}\,\includegraphics[scale=0.5]{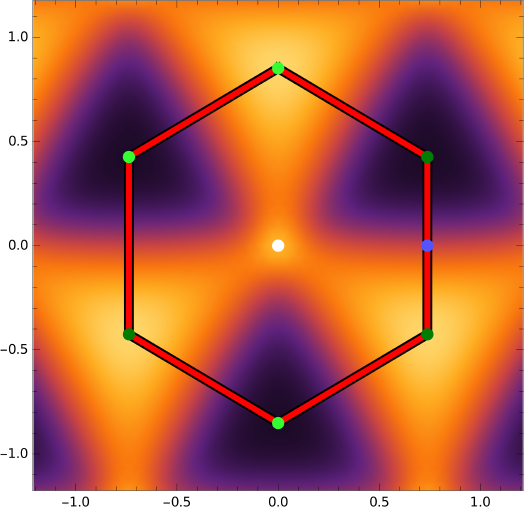}\,\includegraphics[scale=0.5]{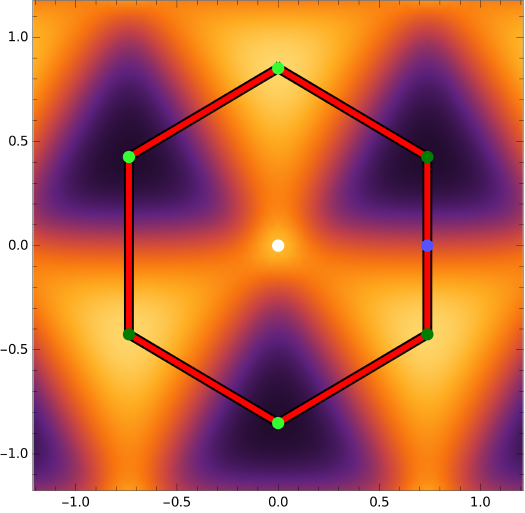}
	
	\centering\includegraphics[scale=0.78]{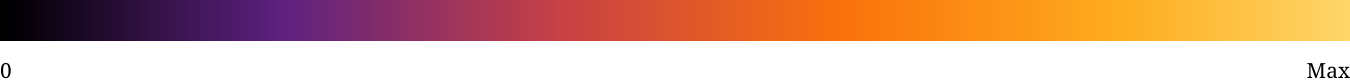}
	\vspace{-0.2cm}\centering\caption{Wave function of the Goldstone magnon (first and second columns, computed via BSE/TDA and BSE/RPA respectively) and the lowest bright exciton (third and fourth columns, computed via BSE/TDA and BSE/RPA respectively) of ferromagnetic graphene for COM momentum $\bm{Q}$ (top to bottom) corresponding to the high symmetry points $\boldsymbol{\Gamma}$, $\bm{M}$, $\bm{K}$. Whenever a state is degenerate (see Fig.~\ref{fig:magnon-exciton}), wave functions of both degenerate states were summed. Highlighted points in the Brillouin zone are $\boldsymbol{\Gamma}$ (black or white, for contrast), $\bm{M}$ (blue), $\bm{K}/\bm{K}'$ (green), with then darker green points representing the sampled points for finite $\bm{Q}$ computations.\label{fig:magnon_wavefunction}}
\end{figure*}

Computing these, {one observes that the two different forms of the BSE yield very similar} wave functions for ferromagnetic graphene, as can be seen in Fig.~\ref{fig:magnon_wavefunction}. However, some differences can still be seen, especially in the Goldstone magnon at $\bm{Q}=0$ where the RPA wave function much more clearly showcases the expected fully flat character of a collective spin flip. The first exciton is essentially indistinguishable, matching the expectations set by the essentially identical optical conductivity between the two forms of the BSE (see Fig.~\ref{fig:optical-response}).

\end{appendices}

	\bibliographystyle{ieeetr}
	\bibliography{magnon2}
\end{document}